\documentclass{aa}

\usepackage{graphicx}
\usepackage{graphicx}
\usepackage{subcaption}
\usepackage{txfonts}
\usepackage{natbib}

\bibpunct{(}{)}{;}{a}{}{,}

\begin{document}

\title{Lensed quasar search via time variability with the HSC transient survey}

\author{Dani C.-Y. Chao\inst{1, 2} \and James H.-H. Chan\inst{3} \and Sherry H. Suyu\inst{1,2,4} \and Naoki Yasuda\inst{5} \and Anupreeta More\inst{5,6} \and Masamune Oguri\inst{5,7,8} \and Tomoki Morokuma\inst{9,5} \and Anton T. Jaelani\inst{10,11}}

\institute{Max Planck Institute for Astrophysics,
              Karl-Schwarzschild-Str. 1, 85741 Garching, Germany\\
              \email{dchao@mpa-garching.mpg.de}
              \and
              Physik-Department, Technische Universität München, 
              James-Franck-Straße 1, 85748 Garching, Germany
              \and
              Institute of Physics, Laboratory of Astrophysics, Ecole Polytechnique Fédérale de Lausanne (EPFL), 
              Observatoire de Sauverny, 1290, Versoix, Switzerland
              \and
              Institute of Astronomy and Astrophysics, Academia Sinica, 
              11F of ASMAB, No.1, Section 4, Roosevelt Road, Taipei 10617, Taiwan
              \and
              Kavli Institute for the Physics and Mathematics of the Universe (WPI), The University of Tokyo Institutes for Advanced Study, The University of Tokyo, 5-1-5 Kashiwanoha, Kashiwa,Chiba 277-8583, Japan
              \and
              The Inter-University Center for Astronomy and Astrophysics, 
              Post bag 4, Ganeshkhind, Pune, 411007, India
              \and
              Department of Physics, University of Tokyo, 
              7-3-1 Hongo, Bunkyo-ku, Tokyo 113-0033, Japan
              \and
              Research Center for the Early Universe, University of Tokyo, 
              7-3-1 Hongo, Bunkyo-ku, Tokyo 113-0033, Japan
              \and
              Institute of Astronomy, Graduate School of Science, The University of Tokyo, 
              2-21-1 Osawa, Mitaka, Tokyo 181-0015, Japan
              \and
              Department of Physics, Kindai University, 
              3-4-1 Kowakae, Higashi-Osaka, Osaka 577-8502, Japan
              \and
              Astronomy Study Program and Bosscha Observatory, FMIPA, Institut Teknologi Bandung, 
              Jl. Ganesha 10, Bandung 40132, Indonesia
              }

\date{Received --}

\abstract
{Gravitationally lensed quasars are useful for studying astrophysics and cosmology, and enlarging the sample size of lensed quasars is important for multiple studies. In this work, we develop a lens search algorithm for four-image (quad) lensed quasars based on their time variability. In the development of the lens search algorithm, we constructed a pipeline simulating multi-epoch images of lensed quasars in cadenced surveys, accounting for quasar variabilities, quasar hosts, lens galaxies, and the point spread function variation. Applying the simulation pipeline to the Hyper Suprime-Cam (HSC) transient survey, an ongoing cadenced survey, we generated HSC-like difference images of the mock lensed quasars from the lens catalog of \citet{OM10}. With the difference images of the mock lensed quasars and the variable objects from the HSC transient survey, we developed a lens search algorithm that picks out variable objects as lensed quasar candidates based on their spatial extent in the difference images. We tested the performance of our lens search algorithm on a sample combining the mock lensed quasars and variable objects from the HSC transient survey. Using difference images from multiple epochs, our lens search algorithm achieves a high true-positive rate (TPR) of 90.1\% and a low false-positive rate (FPR) of 2.3\% for the bright quads (the third brightest image brightness $m_\text{3rd} < 22.0 \text{ mag}$) with wide separation (the largest separation among the multiple image pairs $\theta_\text{LP} > 1.5\arcsec$). With a preselection of the number of blobs in the difference image, we obtain a TPR of 97.6\% and a FPR of 2.6\% for the bright quads with wide separation. Even when difference images are only available in one single epoch, our lens search algorithm can still detect the bright quads with wide separation at high TPR of 97.6\% and low FPR of 2.4\% in the optimal seeing scenario, and at TPR of $\sim$94\% and FPR of $\sim$5\% in typical scenarios. 
Therefore, our lens search algorithm is promising and is applicable to ongoing and upcoming cadenced surveys, particularly the HSC transient survey and the Rubin Observatory Legacy Survey of Space and Time, for finding new lensed quasar systems.} 


\maketitle
%

\section{Introduction}

Gravitationally lensed quasars are powerful tools to study astrophysics and cosmology. For example, they can be used to probe the substructure of dark matter  \citep[e.g.,][]{dm_sub_01, dm_sub_02, dm_sub_03, dm_sub_04, Gilman_Sub}, to examine the formation and evolution of supermassive black holes \citep{high_z_lensed_quasar}, and to study the properties of quasar host galaxies \citep[e.g.,][]{quasar_host_01, quasar_host_02}. With the time delays between multiple images, we can also measure the Hubble constant, $H_0$, which is crucial for testing cosmological models and studying dark energy \citep[e.g.,][]{h0_refsdal,  H0_Birrer, holicow_h0_01, holicow_h0_02}. \\

Given the importance of lensed quasars, there have been several systematic searches of these objects. The Cosmic Lens All Sky Survey \citep[CLASS;][]{class_01, class_02} discovered a large sample of lensed quasars in radio wavelengths by first selecting radio sources that have a compact flat spectrum and then looking for the high-resolution, multiply-imaged components in the radio follow-up. In the optical, the Sloan Digital Sky Survey \citep[SDSS;][]{SDSS_survey} Quasar Lens Search \citep[SQLS;][]{sqls_01, sqls_02, sqls_03, sqls_04, BQLS}, which has found the largest sample of lensed quasars ($\sim$60), started with spectroscopic confirmed SDSS quasars. To these, they applied a morphological selection for potential lens candidates with narrow separation and a color selection for potential lens candidates that are deblended in the SDSS; next, these authors conducted follow-up observations to confirm these candidates after a visual inspection of the objects selected by the morphological or color selection. Both CLASS and SQLS start with spectroscopic searches, and further look into the morphology or images for lensed quasar candidates. In the present stage, further astrophysical or cosmological studies with lensed quasars are still limited by the small sample size of lensed quasars, given the fact that lensed quasars are rare.\\ 

Fortunately, owing to the increased depth and much larger areal coverage, many ongoing and upcoming wide-field surveys provide a great pool of new lensed quasars and allow lens searches directly with images. With the aim of exploiting the unprecedented advantage of those surveys, many dedicated searches have been initiated with new lens search techniques. Some lens search techniques explore multiband catalogs and employ various cuts in magnitude and color spaces to find lensed quasars \citep[e.g.,][]{Adriano_method, Adriano_alone, Strides_method, Peter_Williams_method, Rusu_18}. Some other techniques look for lensed quasars by examining their configuration or recognizing their pattern in images \citep[e.g.,][]{Adriano_method, chitah}. Although not specific to lensed quasars, \textsc{Space Warps} \citep{space_warp_01, space_warp_02} shows that lensed quasars can also be found through citizen science.\\
%

With the exceptional resolution of \textit{Gaia} \citep{Gaia_survey}, it is possible to conduct quasar lens search by looking for multiple detections in \textit{Gaia} or comparing the flux and position offsets from other surveys for objects in SDSS, Panoramic Survey Telescope and Rapid Response System \citep[Pan-STARRS;][]{panstarrs_survey} \citep[e.g.,][]{Cameron_01, Cameron_02, Cameron_03, Ostrovski_gaia_panstarrs_03}, Dark Energy Survey \citep[DES;][]{des_survey} \citep[e.g.,][]{Adriano_gaia_des_01, Adriano_gaia_des_02}, or Kilo-Degree Survey \citep[KiDS;][]{kids_survey} \citep{Spinello_gaia_kids}. Using the ability of \textit{Gaia} to carry out multiple detections to search for lensed quasars could also be applied to the ATLAS\footnote{VST-ATLAS,  The VLT Survey Telescope ATLAS \citep{atlas_survey}} footprint \citep{Adriano_atlas}. We can also combine \textit{Gaia} detections with other lens searches to find lensed quasars. While the search for lensed quasars with small separations benefits most from \textit{Gaia}'s exceptional resolution, \textit{Gaia} usually can detect only one or two multiple image(s) of a lens system, which is limited by its shallow depth, $i \lesssim 20.7$.\\

In the near future, thousands of lensed quasars are expected to be detected by the Rubin Observatory Legacy Survey of Space and Time \citep[LSST;][]{lsst_survey} \citep[][hereafter OM10]{OM10}. With further imaging and spectroscopic analysis, we are able to discover and confirm these LSST detections, enlarging the sample size of lensed quasars by at least an order of magnitude. Moreover, the LSST is a cadenced survey that will reveal the time variability of variable objects, such as lensed quasars with significant variability\footnote{There are known lensed quasars with weak intrinsic variability, which would be difficult to find through their variability.}. In a cadenced survey, difference imaging is helpful in eliminating non-variable objects. As first suggested by \citet{Kochanek_method}, looking for extended objects in the difference images is an effective means to find lensed quasars, since almost all the other variable objects (non-lensed variable objects) are point sources. Therefore, a method based on time variability benefits lens searches given the enormous number of objects in the LSST.\\ 

Inspired by \citet{Kochanek_method}, we developed a lens search algorithm for lensed quasars based on their time variability. The Hyper Suprime-Cam \citep[HSC;][]{Miyazaki_hsc, Aihara_SSP} transient survey \citep{hsc_transient} provides a great opportunity to simulate the difference images that could be used for the development of the lens search algorithm and to test the lens search performance. The HSC transient survey is an ongoing cadenced survey of the Subaru Telescope \citep{Miyazaki_subaru} with similar image quality expected for the LSST. \\ 

In this work, we build a simulation pipeline for producing time-varied images of lensed quasars, and apply the simulation pipeline to the HSC transient survey to generate HSC-like difference images of mock lensed quasars. With the HSC-like difference images of lensed quasars, we develop a lens search algorithm that picks out variable objects with large spatial extent in the difference images and classifies these objects as lensed quasar candidates. We further test the performance of our lens search algorithm in the HSC transient survey. Although the simulation and the search algorithm could also be applied to lenses with two-image configurations (double), we focus on lenses with four-image configuration (quad). In spite of the higher number of doubles than quads, quads provide more constraints on the lens potential distribution and stellar mass fraction. Furthermore, the configuration of quads is so unique that it is unlikely to be mistakenly identified, while many objects such as quasar binaries are easily mistaken as doubles.\footnote{The double used in \citet{H0_Birrer} for time-delay cosmography has a special configuration in which its host galaxy is quadruply lensed; this makes it possible to perform an analysis similar to quadruply lensed quasars.}\\

The organization of the paper is as follows. In Sec.~\ref{sec:simulation}, we present a new simulation pipeline of time-varied lensed images, and apply the new simulation pipeline to the HSC transient survey in Sec.~\ref{sec:application}. The lens search algorithm is detailed in Sec.~\ref{sec:method}. The lens search performance is shown in Sec.~\ref{sec:performance} and we conclude in Sec.~\ref{sec:conclusion}.

\section{Simulation of time-varying lensed quasars images}
\label{sec:simulation}
\begin{figure*}
\centering
\includegraphics[scale=0.85]{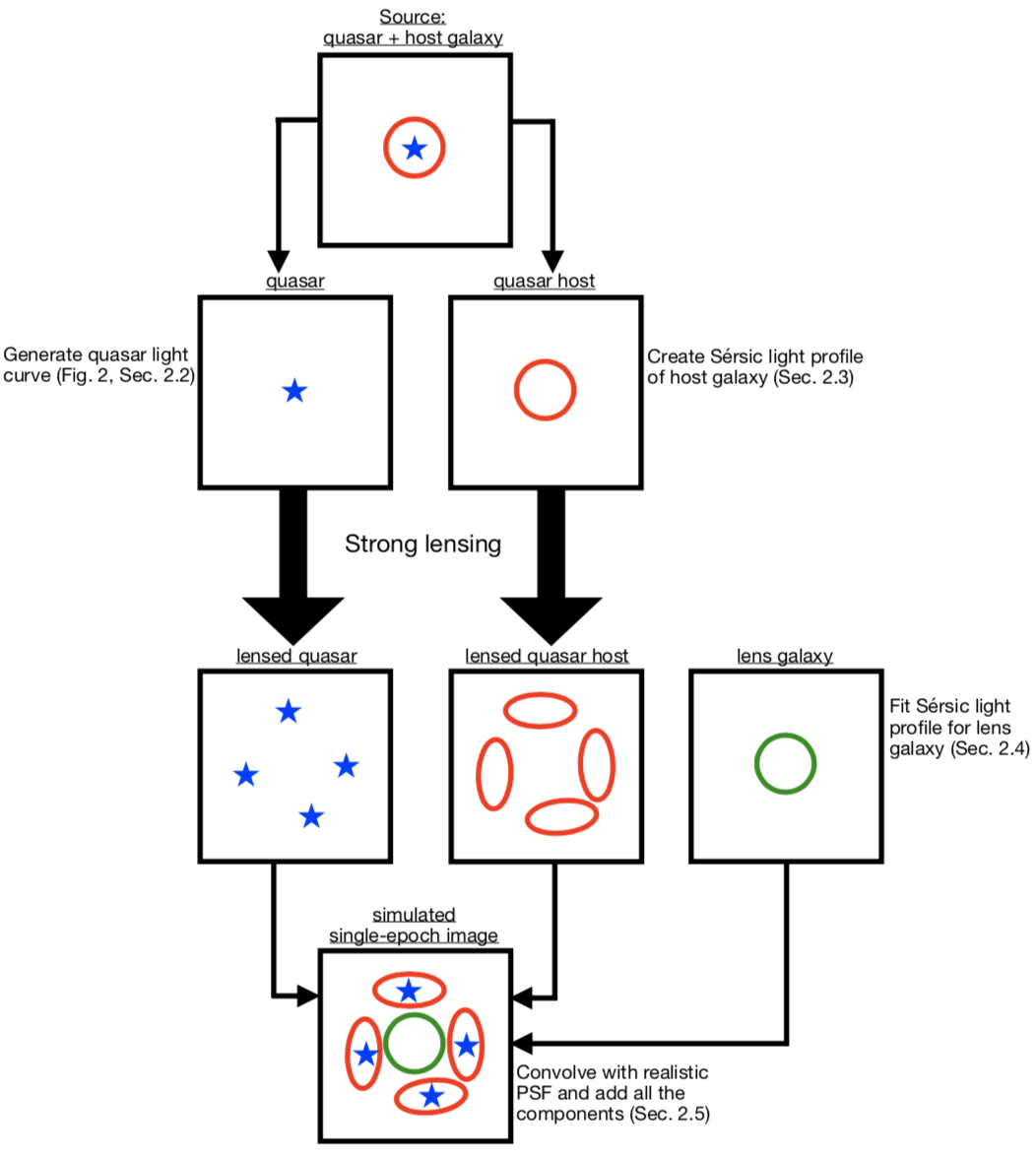}
\caption{Flow chart illustrating the simulation of time-varying lensed quasars images. For quasars, mock quasar light curves are generated based on their source redshifts $z_\text{s}$, and being shifted in time and magnified the mock quasar light curves on the image plane (see Fig.~\ref{fig:quasar_light_ex}). For quasar hosts, their S\'ersic light profiles are created on the source plane and their lensed images are produced on the image plane assuming a SIE model. Lens galaxies are selected from a real survey based on their lens redshifts $z_\text{d}$ and velocity dispersions $\sigma$, fitting the selected galaxies with S\'ersic light profiles, and using the fitted S\'ersic light profiles to produce their images on the image plane. Once all the components for one lens system (image of lensed quasar, image of lensed quasar host, and image of lens galaxy) are produced, these components are convolved with a realistic PSF and are added to obtain one complete image.} 
\label{fig:overview}
\end{figure*}

We present a simulation pipeline for creating realistic images of mock lensed quasars in a series of time.  The mock images produced by this pipeline are useful for developing a lens search algorithm based on time variability. We used the mock configuration from the catalog created by OM10 to generate the simulated lensed quasars. In order to have realistic simulations, we simulated not only the light from the variable object - the lensed quasar in this work - but also the light from the (lensed) host galaxy and lens galaxy. This also takes into account possible residuals on the difference image due to potential imperfection of the difference imaging method. The flow chart of this simulation pipeline is shown in Fig.~\ref{fig:overview}.\\

\subsection{Mock lenses from OM10}
\citet{OM10} (OM10) predicted the distribution of lensed quasars and produced a mock catalog of lensed quasars. For each lens system, OM10 provides the source position $(\eta_1, \eta_2)$, source redshift $z_\text{s}$, unlensed \textit{i}-band magnitude $m_\text{s}$ of the quasar, lens redshift $z_\text{d}$, velocity dispersion $\sigma$, ellipticity $e$, and position angle (PA) $\theta_e$ of the lens galaxy. Also, OM10 provides the lensed image positions, the magnifications, and the time delays calculated by \textsc{glafic} \citep{Masamune_glafic}, with the assumption of a singular isothermal ellipsoid (SIE) model \citep{SIE} for the lens and an external shear accounting for the effect of the lens environment \citep[e.g.][]{Shear_01, Shear_02, Shear_03}. The convergence $\kappa$ from the SIE model of the lens is given by
\begin{equation}
\label{eq:sie}
\kappa(\theta_1, \theta_2) = \frac{\theta_\text{Ein} \sqrt{1-e}}{2} \frac{\lambda(e)}{\sqrt{\theta_1^2+(1-e)^2 \theta_2^2}},
\end{equation}
where $(\theta_1, \theta_2)$ is the coordinate of the lens, $\theta_\text{Ein}$ is the Einstein radius in arcsec, $e$ is the ellipticity, and $\lambda(e)$ is the dynamical normalization defined in \citet{Masamune_dyn_norm}. The lens potential of the external shear $\phi$ is given by
\begin{equation}
\label{eq:shear}
\phi(\theta_1, \theta_2) = \frac{\gamma}{2}(\theta_1^2 - \theta_2^2)\cos 2\theta_\gamma + \gamma \theta_1 \theta_2 \sin 2\theta_\gamma,
\end{equation}
where $\gamma$ is the magnitude of the external shear and $\theta_\gamma$ is the orientation of the external shear.

\subsection{Quasar}
\label{sec:quasar}
For the point sources, that is, the lensed quasars without the (lensed) host galaxies, we used the source redshift $z_\text{s}$, unlensed \textit{i}-band magnitude $m_\text{s}$, lensed image positions, magnifications, and time delays provided by OM10 to generate the simulated image for each lensed quasar. \\

We generated the source light curve of quasars via \textsc{carma\_pack} \citep{Kelly_2009,  Kelly_2014} based on $z_\text{s}$ and $m_\text{s}$ from OM10. The \textsc{carma\_pack} package was developed to quantify stochastic variability, especially for quasar variability, and the quasar light curves are generated with a damped-random-walk model based on magnitude $m$ and the fitting relation on redshifts $z$ in \citet{Kelly_2009}. We then shifted (in time) and magnified the generated source light curve according to the time delay and magnification of each lensed image.\footnote{We did not explicitly include microlensing variability. The lensed quasar variability is changed but not canceled out by microlensing variability, so lensed quasars still appear as multiple point-like variable objects in the difference images even when there is a microlensing effect.}\\ 

\begin{figure*}
\includegraphics[width=\textwidth]{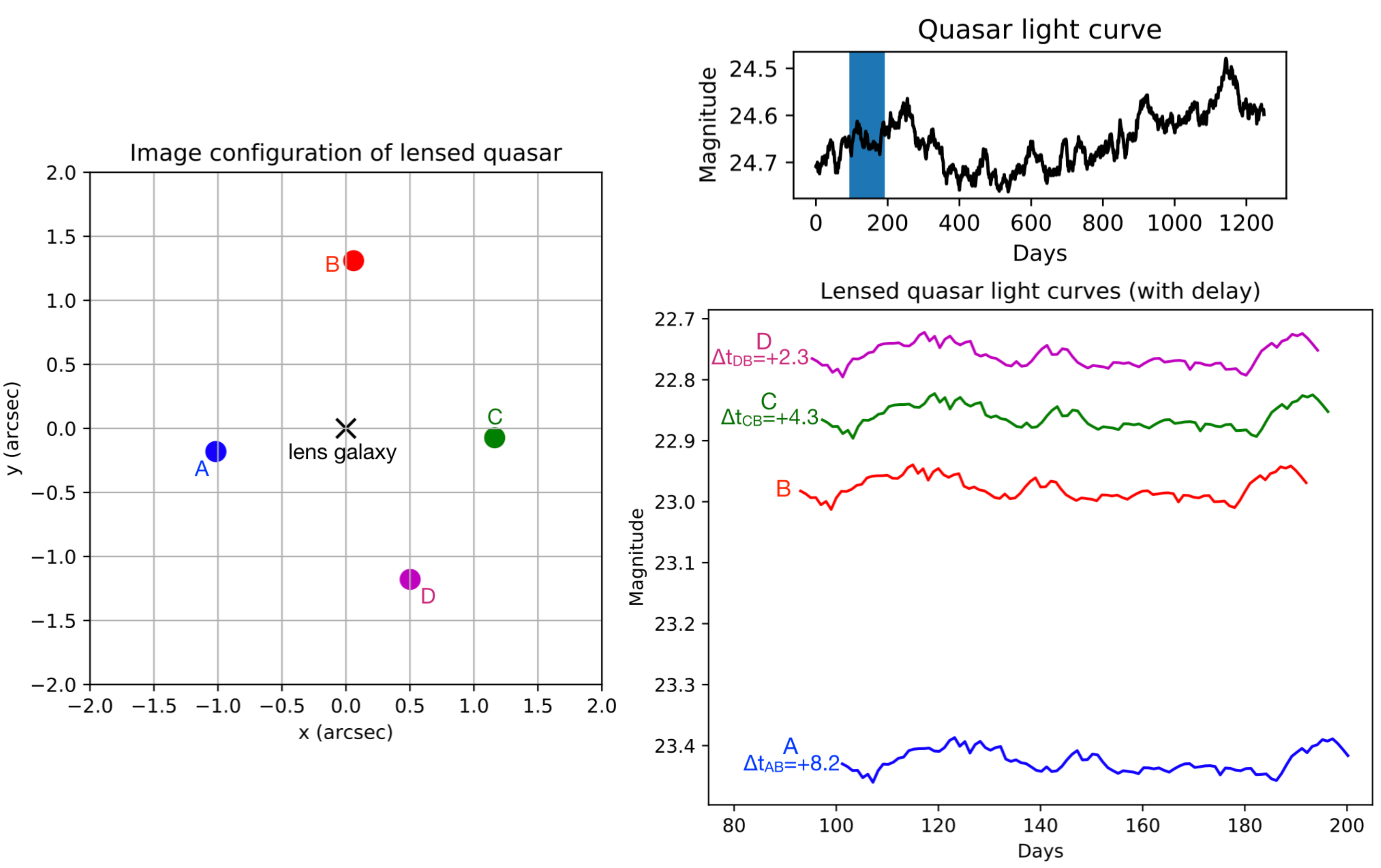}
\caption{Illustration of generating light curves for a lensed quasar without its (lensed) host galaxy. Left: Positions of the four multiple images (A, B, C, and D) and the lens galaxy (cross). Top right: Mock quasar light curve. Bottom right: The lensed light curves of the four multiple images in the left panel, for the period marked by the blue band in the top right panel. } 
\label{fig:quasar_light_ex}
\end{figure*}

Fig.~\ref{fig:quasar_light_ex} is an illustration of the procedure to generate light curves for lensed quasars without the (lensed) host galaxies. In the left panel, we show the positions of the four multiple images of a symmetry quad lensed by a galaxy at $z_\text{d}=0.23$ from OM10 with source redshift $z_\text{s}=3.26$. The four multiple image positions, A, B, C, and D in the left panel, are relative to the lens galaxy, which is at the center for each lens system in OM10, $(x, y) = (0,0)$. Their magnifications, $|\mu_\text{A}|, |\mu_\text{B}|, |\mu_\text{C}|,$ and $|\mu_\text{D}|$, and the time delays relative to Image B, $\Delta t_\text{AB}, \Delta t_\text{CB},$ and $\Delta t_\text{DB}$, are listed in Table~\ref{tab:quasar_ex} (where $\Delta t_\text{XB} = t_\text{X} - t_\text{B}$). The source light curve that we generated with \textsc{carma\_pack} based on the source redshift $z_\text{s}=3.26$ is shown in the top right panel of Fig.~\ref{fig:quasar_light_ex}. The bottom right panel of Fig.~\ref{fig:quasar_light_ex} shows the light curves of the four multiple images (A, B, C, and D in the left panel), which are produced by shifting and magnifying part of the source light curve in the top right panel (blue band) with the corresponding time delays and magnifications in Table~\ref{tab:quasar_ex}.\\

\begin{table}
\centering
\begin{tabular}{c c c}
\hline
Image & Magnification ($|\mu|$) & Time delay ($\Delta t$)  \\
 &  & (in days relative to Image B)\\
\hline
A  & 3.1 & +8.2\\
B & 4.7 & 0\\
C & 5.2 & +4.3\\
D & 5.7 & +2.3\\
\hline
\end{tabular}
\caption{Magnifications and time delays of the lensed quasar in Fig.~\ref{fig:quasar_light_ex}.}
\label{tab:quasar_ex}
\end{table}

\subsection{Quasar host galaxy}
\label{sec:quasar_host}
Since OM10 only provides the configuration of point sources and there is no definite morphological relation between quasar and the host galaxy with little scatter, we randomly drew the morphological properties based on the S\'ersic profile for the host galaxy of quasar and used \textsc{glafic} \citep{Masamune_glafic} to generate the lensed image of host galaxy. We conservatively adopted broad ranges for the parameters in the S\'ersic profile of the host galaxy \citep[based on, e.g.,][]{host_galaxy_prop, P06, P15, B10}. The thresholds of the host galaxy properties used in generating the lensed images are listed in Table~\ref{tab:host_properties}.\\  

\begin{table}
\centering
\begin{tabular}{c c c}
\hline
Property & Min. & Max. \\
\hline
log $L_{\text{host}}$ ($L_{\sun}$) & 9 & 13\\
$e$ & 0 & 0.8\\
$\theta_e$ (deg) & 0 & 180\\
$r_\text{e}$ (kpc) & 1 & 10\\
$n$ & 1 & 5\\
\hline
\end{tabular}
\caption{Ranges of the quasar host galaxy properties - luminosity $L_{\text{host}}$, ellipticity $e$, PA $\theta_e$, effective radius $r_\text{e}$, and S\'ersic index $n$.}
\label{tab:host_properties}
\end{table}

We put the host galaxy with its random S\'ersic profile at the same source position as the (unlensed) quasar provided by OM10, and we adopted the same SIE model, external shear, and cosmology from OM10 for \textsc{glafic} to generate the lensed image of the host galaxy. Therefore, the lensed host galaxy image positions are consistent with the lensed quasar image positions.\\

\subsection{Lens galaxy}
\label{sec:lens}
For the lens galaxies, we assigned the light profile to the lens galaxy for each lens system with fitted profile of galaxy from a real survey because only the mass model of lens galaxy is given by OM10. We denoted such a real survey "$S_\text{imag}$", where $S_\text{imag}$ can be any survey devoted to finding lens systems, such as the SDSS or the HSC survey.\\

We began with a catalog created by cross-matching the galaxies from $S_\text{imag}$ with another survey "$S_\text{spec}$", which provides spectroscopic redshift $z_\text{spec}$ and velocity dispersion $\sigma_\text{spec}$. We note that if $S_\text{imag}$ itself has information about spectroscopic redshift and velocity dispersion, we do not have to cross-match $S_\text{imag}$ with another survey. From the cross-matched catalog, we selected "matched galaxies" for each lens galaxy with the closest values to the lens redshift $z_{\text{d}, \text{OM10}}$ and the velocity dispersion $\sigma_{\text{OM10}}$ from the mass model provided by OM10. We first selected the galaxies with
\begin{equation}
\label{eq:lens_select_z}
|d_z| < 0.01
\end{equation}
and
\begin{equation}
\label{eq:lens_select_sigma}
|d_{\sigma}| < \Delta \sigma,
\end{equation}
where $d_z = z_{\text{spec}}-z_{\text{d}, \text{OM10}}$, $d_{\sigma} = \sigma_{\text{spec}}-\sigma_{\text{OM10}}$, and $\Delta \sigma$ is the error in $\sigma_{\text{spec}}$ from $S_\text{spec}$. We then got the matched galaxies from the selected galaxies with the smallest values of $d_{z,\sigma}$, where $d_{z,\sigma}$ is defined as
\begin{equation}
d_{z, \sigma} = \sqrt{d_z^2 + d_{\sigma}^2}.
\end{equation}
Ideally, for one lens galaxy in OM10, we could have ten galaxies  with the smallest $d_{z, \sigma}$ from the cross-matched catalog as matched galaxies for the lens light. However, for some lens systems in OM10, the number of selected galaxies from Eqs.~\ref{eq:lens_select_z} and \ref{eq:lens_select_sigma} is not enough to have ten matched galaxies. For such cases, we took all the available galaxies passing the selection criteria (Eqs.~\ref{eq:lens_select_z} and \ref{eq:lens_select_sigma}) as matched galaxies. In this work, we used the data from the SDSS Data Release 14 \citep[DR14;][]{sdss_14} as $S_\text{spec}$.\\ 

The velocity dispersion $\sigma_\text{SDSS}$ is limited in the cross-matched catalog when $z_{\text{SDSS}} > 1.04$, so we applied only Eq.~\ref{eq:lens_select_z} to select matched galaxies for the lens galaxy with $1.04 < z_{\text{d}, \text{OM10}} < 1.57$; we selected up to ten galaxies as the matched galaxies with the smallest values of $d_z$. Owing to the lack of objects with $z_{\text{SDSS}} > 1.57$ in the cross-matched catalog, we did not simulate the lens systems with $z_{\text{d}, \text{OM10}} > 1.57$. \\

We then obtained the image for each matched galaxy from $S_\text{imag}$, and fit the light distribution of the matched galaxy with the S\'ersic profile. For the fitting, we define 
\begin{equation}
\chi^2 = \sum\limits_{i}^{N_{\text{p}}} \frac{(I_{i, \text{fit}} - I_{i, \text{data}})^2}{\sigma_{i, \text{data}}^2},
\end{equation}
and the reduced $\chi^2$ as 
\begin{equation}
\chi^2_\text{reduced} = \frac{\chi^2}{N_{\text{p}}},
\end{equation}
where $N_{\text{p}}$ is the number of pixels used in fitting, $I_{i, \text{fit}}$ is the fitted intensity of light, $I_{i, \text{data}}$ is the observed intensity of light from $S_\text{imag}$, and $\sigma_{i, \text{data}}$ is the noise from $S_\text{imag}$. We ranked the matched galaxies by $\chi^2_\text{reduced}$. The smaller $\chi^2_\text{reduced}$, the higher is the rank.\\ 

After we had the fitted light profile for each matched galaxy, we compared the ellipticity from the fitted light profile, $e_{\text{fit}}$, and the ellipticity from the mass model given by OM10, $e_{\text{OM10}}$, according to the ranking order. For one matched galaxy, if $|e_{\text{fit}} - e_{\text{OM10}}| < 0.2$, we picked the matched galaxy as the lens galaxy; otherwise, we checked $e_{\text{fit}}$ of the \text{matched galaxy} in the next lower rank. For the lens galaxy that has no matched galaxy with the satisfied value of $e_{\text{fit}}$, we picked the matched galaxy with the smallest value of $\chi^2_\text{reduced}$. In total, we simulated 2033 quads with lens redshift $z_{\text{d}, \text{OM10}} < 1.57$ from OM10.\footnote{The properties of the quasar and lens galaxy used in this paper are available at https://github.com/danichao/OM10-quads-Z\_d-1.57-}\\ 

\subsection{Simulated image with all components}
\label{sec:all_comp}
Once we had the individual simulated parts of a whole lens system (the lensed quasar, lensed host galaxy, and lens galaxy), we convolved these individual parts with the point spread functions (PSFs) in $S_\text{imag}$ for each epoch and added the parts to become one complete image. 
The complete image in an epoch $t$ is expressed as
\begin{equation}
I_t(x, y) = \big (I_{\text{q}, t}(x, y) + I_{\text{h}}(x, y) + I_{\text{d}}(x, y) \big ) \otimes \text{PSF}_t(x, y),
\end{equation} 
where $I_t(x, y)$ is the light distribution at position $(x, y)$ of the complete image, $I_{\text{q}, t}(x, y)$ is the light distribution of lensed quasar, $I_{\text{h}}(x, y)$ is the light distribution of (lensed) host galaxy, $I_{\text{d}}(x, y)$ is the light distribution of lens galaxy, $\text{PSF}_{t}(x, y)$ is the PSF from $S_\text{imag}$, and $\otimes$ represents the convolution.  After we generated complete images for all the epochs in $S_\text{imag}$, we obtained the images of the mock lensed quasar in the time series of $S_\text{imag}$.

\section{Application to the Hyper Suprime-Cam Survey}
\label{sec:application}
In this section, we demonstrate our simulation in the HSC transient survey, and describe how we generate realistic difference images that will be used to develop the search algorithm through the pipeline of the HSC transient survey.\\

\subsection{HSC transient survey}
\label{sec:HSC}
The HSC transient survey was done in the COSMOS \citep{COSMOS} field as part of the HSC-SSP (Subaru Strategic Program; \citealt{Aihara_SSP, HSC_Miyazaki_2018,HSC_Komiyama_2018, HSC_Kawanomoto_2018, HSC_Furusawa_2018}) observation from November 2016 to April 2017 with a pixel size of 0.168 arcsec, covering 1.77 $ \deg^{2}$ in the Ultra-Deep layer to a median depth per epoch of 25.9 mag in \textit{i} band. Since OM10 has only \textit{i}-band magnitude, we simulated lenses only in \textit{i} band. The HSC transient survey has 13 epochs (nights of observations) over six months for \textit{i} band (Table~\ref{tab:HSC_epochs}).\\
\begin{table}
\centering
\begin{tabular}{ c c }
\hline
Observation date & Seeing (arcsec) \\ 
\hline          
2016-11-25 & 0.83 \\
2016-11-29 & 1.16\\
2016-12-25 & 1.25\\
2017-01-02 & 0.68\\
2017-01-23 & 0.70\\
2017-01-30 & 0.76\\
2017-02-02 & 0.48\\
2017-02-25 & 0.72\\
2017-03-04 & 0.69\\
2017-03-23 & 0.66\\
2017-03-30 & 0.98\\
2017-04-26 & 1.24\\
2017-04-27 & 0.58\\            
\hline
\end{tabular}
\caption{Observation dates and seeing in \textit{i} band from HSC transient survey.}
\label{tab:HSC_epochs}
\end{table}

To obtain the difference image for each epoch, we need the reference image to subtract from the single-epoch image. In the HSC transient survey, the reference images are the deep reference images produced by co-adding multiple exposures from multiple epochs during March 2014 to April 2016 (Table~\ref{tab:ref_epochs}). Only the exposures with seeing better than 0.7 arcsec were used to create the deep reference images.\\ 

The HSC transient survey uses the method in \citet{Robert_diff} and \citet{Alard_diff} for the difference imaging. The single-epoch image is created by co-adding several warped images, and each warped image corresponds to a distortion-corrected image of the sky from a single exposure. For each epoch, the difference imaging was performed on every warped image by subtracting the deep reference image to produce the warped difference images, and all the warped difference images were then co-added to create the deep difference image. The HSC transient survey convolves the exposures used in creating the deep reference image (Table~\ref{tab:ref_epochs}) with a kernel to match the PSF of each warped image when performing the difference imaging process.\\  

To generate realistic deep difference image of the mock lenses, we selected the locations that have no object within 5 arcsec, which we call "empty regions", and we injected our mock lenses into these empty regions with the HSC pipeline \citep{Bosch_pip} version 4.0.5, for both the warped images in each epoch and the deep reference image. Once the injection was done, we used the difference imaging method mentioned above to create the deep difference image of the mock lenses. Practically, we selected 75 empty regions in the "patch" of sky identified with the HSC $\texttt{Tract} = 9813, \texttt{Patch} = (3,4)$ to inject the mock lenses. In this patch, 44 warped images from five epochs were used to create the deep reference image (Table~\ref{tab:ref_epochs}), and 175 warped images were used to create the deep difference images for all the 13 epochs in \textit{i} band.\\

\begin{table}
\centering
\begin{tabular}{c  c  c}
\hline
Observation date & MJD & Number of exposures  \\
\hline
2014-03-28 & 56744.4687 & 9\\
                   & 56744.4723 &\\
                   & 56744.4759 &\\
                   & 56744.4874 &\\
                   & 56744.4911 &\\
                   & 56744.4950 &\\
                   & 56744.4987 &\\
                   & 56744.5024 &\\
                   & 56744.5061 &\\
\hline
2015-01-21 & 57043.5571 & 16\\           
                   &  57043.5609 &\\
                   &  57043.5648 &\\
                   &  57043.5738 &\\
                   & 57043.5780 &\\
                   & 57043.5859 &\\
                   & 57043.5898 &\\
                   & 57043.5937 &\\
                   & 57043.6011 &\\
                   & 57043.6047 &\\
                   & 57043.6118 &\\
                   & 57043.6157 &\\
                   & 57043.6195 &\\
                   & 57043.6269 &\\
                   & 57043.6306 &\\
                   & 57043.6455 &\\
\hline
2015-03-20 & 57101.3403 &5\\          
                   & 57101.3489 &\\
                   & 57101.3562 &\\
                   & 57101.3854 &\\
                   & 57101.3925 &\\
\hline 
2015-05-21 & 57163.2518 &12\\           
                   & 57163.2556 &\\
                   & 57163.2595 &\\
                   & 57163.2633 &\\ 
                   & 57163.2672 &\\
                   & 57163.2710 &\\
                   & 57163.2749 &\\
                   & 57163.2787 &\\ 
                   & 57163.2826 &\\ 
                   & 57163.2864 &\\ 
                   & 57163.2903 &\\ 
                   & 57163.2941 &\\ 
\hline
2016-03-04 & 57451.4246 &2\\
                   & 57451.4645 &\\ 
\hline
\end{tabular}
\caption{Exposures used to create the deep reference image for the HSC patch ID, $\texttt{Tract} = 9813, \texttt{Patch} = (3,4)$, in \textit{i} band.}
\label{tab:ref_epochs}
\end{table}

\subsection{Convolution with the HSC PSFs and lens injection}
In order to obtain realistic deep difference image, we injected the mock lens into both the warped image and the deep reference image. For the warped images, the difference imaging method is sensitive to every warped image used to create the deep difference image for each epoch, so accounting for the variations in the PSFs of the warped images is crucial. Meanwhile, the deep reference image used for the subtraction is co-added, so weighting the variability from the dates in Table~\ref{tab:ref_epochs} plays a more important role. We detail the injection methods for the warped image and the deep reference image below.\\   

For the lens injection into the warped images of one epoch, we first simulated the images for the whole lens system as mentioned in Sec.~\ref{sec:simulation}, accounting for the quasar variability in the epoch, and then we created the complete image by convolving with the PSFs of the empty region where the lens system is injected into the warped images. Since one epoch is composed of several warped images, the PSF of the same empty region changes across multiple warped images. Therefore, for one lens system in one epoch, the complete images are generated by convolving with different PSFs for different warped images, and we further injected the complete images into the corresponding warped images.\\

For the lens injection into the deep reference image, we first weighted the images of the mock lens on an epoch basis by the number of the warped images used in one epoch, and unlike the lens injection into the warped images, we convolved the weighted image with the PSF from the deep reference image and further injected the mock lens directly into the deep reference image. In the deep reference image, the total light distribution at position $(x, y)$, $I_\text{ref}(x, y)$, can be described as  
\begin{equation}
I_\text{ref}(x, y) = \Big ( \sum\limits_t \frac{I_{\text{q}, t}(x, y) \cdot N_t}{N_\text{total}} + I_\text{h}(x, y) + I_\text{d}(x, y) \Big ) \otimes \text{PSF}_\text{ref} (x, y),
\end{equation}
where $t =$ 2014-03-28, ..., 2016-03-04 (Table~\ref{tab:ref_epochs}), $I_{\text{q}, t}(x, y)$ is the light distribution of lensed quasar in $t$, $N_t$ is the number of warped images used in $t$, $N_\text{total}$ is the total number of warped images ($N_\text{total} = 44$ in our work), $I_\text{h}(x, y)$ is the light distribution of (lensed) quasar host galaxy, $I_\text{d}(x, y)$ is the light distribution of lens galaxy, $\text{PSF}_\text{ref} (x, y)$ is the PSF from the deep reference image, and $\otimes$ represents the convolution. Injecting mock lenses into the deep reference image via this method takes into account the variability of lensed quasar across the multiple epochs used in the deep reference image.\\

\begin{figure*}
\includegraphics[width=\textwidth]{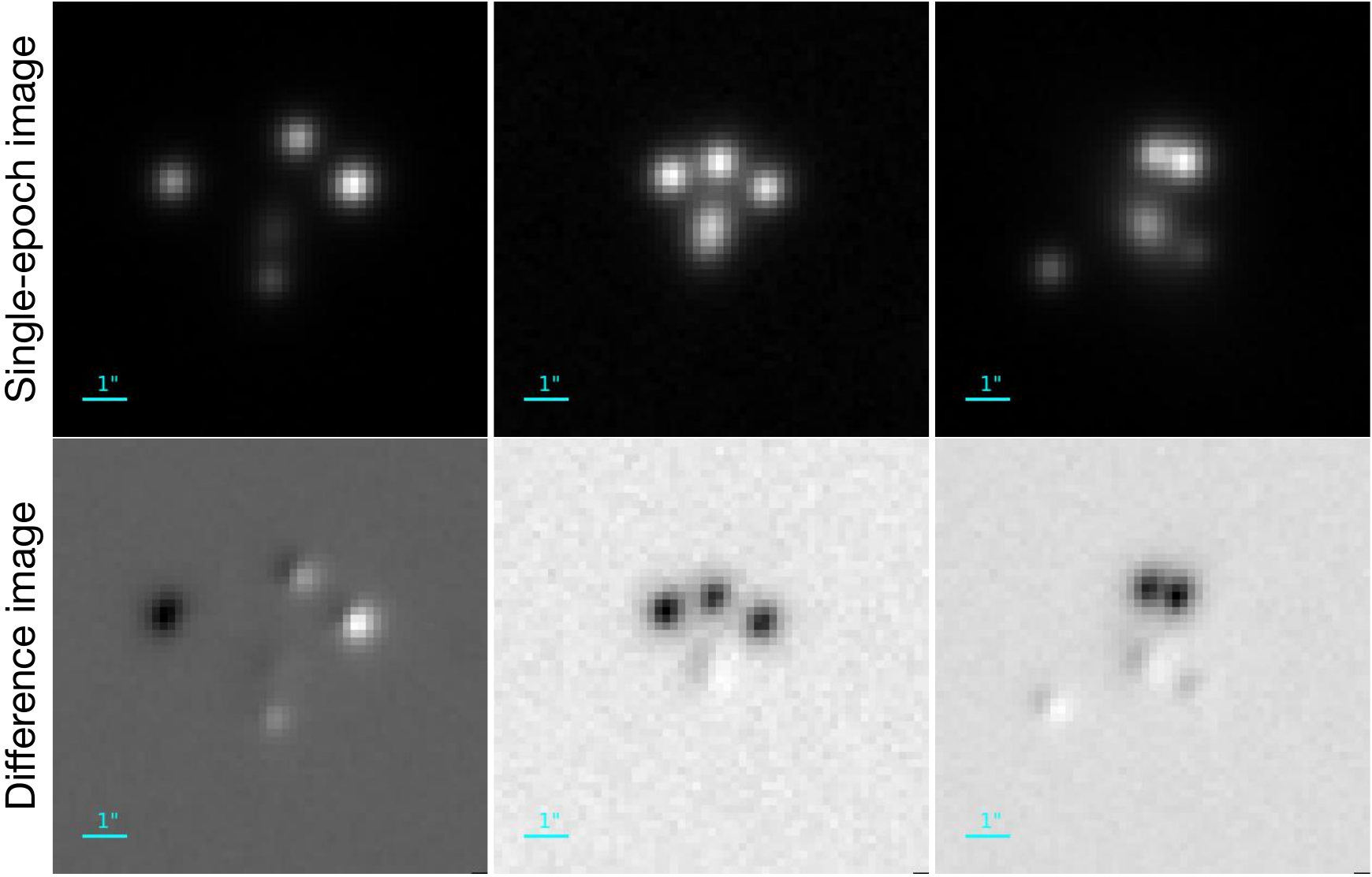}
\caption{Examples of lenses injected into the HSC transient survey. Top: Single-epoch images co-added by the pipeline of the HSC transient survey. Bottom: Corresponding difference images of the top row produced by the image subtraction in the pipeline of the HSC transient survey. The size of each image cutout is $10\arcsec \times 10\arcsec$.}
\label{fig:lens_exs}
\end{figure*}

We simulated the complete image for 2033 quads ($z_{\text{d}, \text{OM10}} < 1.57$) from OM10, and the size of each complete image is $10\arcsec\times10\arcsec$ with the lens at the center (top row in Fig.~\ref{fig:lens_exs}). The quads are injected into the empty region randomly, and the empty regions are used repeatedly for different quads. For simplicity, we did not add Poisson noise associated with the mock lens systems when injecting the quads into the empty region, given the small impact of the Poisson noise: We find that the sky background noise generally dominates over Poisson noise in single-epoch images, although the addition of Poisson noise would slightly affect the pixel counts for some bright or wide-separation lens systems (see Sec.\ref{sec:performance_basic}).  We defer the inclusion of Poisson noise, and the propagation of noise from the single-epoch images into the difference images to future work. After the injection of the complete images, we applied the deep difference imaging method to obtain the deep difference image for the mock quads (bottom row in Fig.~\ref{fig:lens_exs}). As mentioned in Sec.~\ref{sec:HSC}, we first get the warped difference images of injected lenses as intermediate products, and the deep difference images are created by co-adding the warped difference images. The top row in Fig.~\ref{fig:lens_exs} are examples of the single-epoch images we obtain after the lens injection processed by the pipeline of the HSC transient survey, and we can see the light from both the lensed quasar and lens galaxy in each panel, which appear as "bright spots". The bottom row in Fig.~\ref{fig:lens_exs} are the corresponding deep difference images of the top row, and the "dark spots" (bright spots) are the spots that become fainter (brighter) than they were in the reference image. In contrast to the single-epoch images where we can only see bright spots and thus brightness changes are not obvious, brightness changes are clearly visible in the deep difference images through the dark spots (decreasing brightness) and the bright spots (increasing brightness). There are some residuals at the locations of the lens galaxies in the bottom panels of Fig.~\ref{fig:lens_exs}, which confirm that simulating the light from all the components (quasar, host galaxy, and lens galaxy) instead of from only variable object (quasar) is important for realistic mock images because the pipeline might not perform perfect subtraction for all the non-variable components.\\  

\section{Search method: Spatial extent in difference image}
\label{sec:method}
With the multiple-image feature and the variable brightness, strongly lensed quasars are expected to exhibit multiple point-like image residuals in the difference image shown in Fig.~\ref{fig:lens_exs}. However, lensed quasars are often not deblended by the survey pipelines, and most of these show extended morphology. Since most astrophysical variable sources are isolated and point-like (e.g., variable stars, unlensed quasars, unlensed supernovae), targeting spatially extended or multiple point-like variable sources in the difference image is an effective approach to find lensed quasars, as previously noted by Kochanek et al. (2006). We developed a lens search algorithm for lensed quasar via a method that quantifies the extendedness of an object in the difference image and selects objects that are large in spatial extent as lens candidates. We present the method that is based on the extendedness in Sec.~\ref{sec:extendedness_only}, and an enhancement of the method with a secondary criterion in Sec.~\ref{sec:extrema_selection}.\\


\subsection{Quantification of the extendedness}
\label{sec:extendedness_only}
The lens search algorithm starts from "HSC variables" in the HSC transient survey. In this work, we define a HSC variable as an object properly detected on the deep difference images at least twice in the HSC transient survey; the detections could be from two different epochs or two different bands. For each HSC variable, we first collected its deep difference images from all the 13 epochs (Table~\ref{tab:HSC_epochs}) and created the "3$\sigma$ mask" for each epoch by picking out the pixels with value larger than $3\sigma$ or smaller than $-3\sigma$. The pixel values in the 3$\sigma$ mask, $I_\text{mask}(i, j)$, are defined as
\begin{equation}
\label{eq:3sigma_mask}
I_\text{mask}(i, j) = 
\begin{cases}
1, & \text{if } |I(i, j)| > 3\sigma(i, j)\\
0, & \text{otherwise}
\end{cases} 
\end{equation}
where $i=1, ..., N_x$ and $j=1, ..., N_y$ are the pixel indices in both the deep difference image and the 3$\sigma$-mask of dimensions $N_x \times N_y$ ($N_x=N_y=59$ in this work), $I(i, j)$ are the pixel values in the deep difference image, and $\sigma(i, j)$ are the estimated 1$\sigma$ uncertainties in the difference image from the HSC transient survey. Figs.~4a and 4b are examples of the deep difference image and the 3$\sigma$ mask from a HSC variable in an epoch, respectively. As indicated in Fig.~4b, 3$\sigma$ mask has several noise peaks in the outskirts that are not related to the HSC variable. Those noise peaks should not be counted in the extendedness of the HSC variable, so we further define the "effective region" by
\begin{equation}
\label{eq:eff_mask}
I_\text{eff}(i, j) = 
\begin{cases}
1, & \text{if } \displaystyle\sum_{i'=i-1}^{i+1} \sum_{j'=j-1}^{j+1} I_\text{mask}(i', j') > 2\\
0, & \text{otherwise}
\end{cases} 
\end{equation}
where $I_\text{eff}(i, j)$ is the pixel value of $(i, j)$ in the effective region. Fig.~4c shows the effective region of the HSC variable in Fig.~4a. Once we have the effective region, we determine the area of the effective region, $A_\text{eff}$, by the sum of the pixel values in the effective region,
\begin{equation}
\label{eq:region_area}
A_\text{eff} =  \sum\limits_{i=1}^{N_x} \sum\limits_{j=1}^{N_y} I_\text{eff}(i, j),
\end{equation}
where $i$ runs from 1 to $N_x$ and $j$ runs from 1 to $N_y$. This area is equivalent to the number of pixels in the effective region, where $ I_\text{eff}=1$. As a result, we evaluate the extendedness of a HSC variable in the deep difference image for a given epoch by the area of this effective region. Doing so, we can discard the effect from the noise peaks in the outskirts and quantify the extendedness of a HSC variable in the difference image.\\

\begin{figure*}
\includegraphics[width=\textwidth]{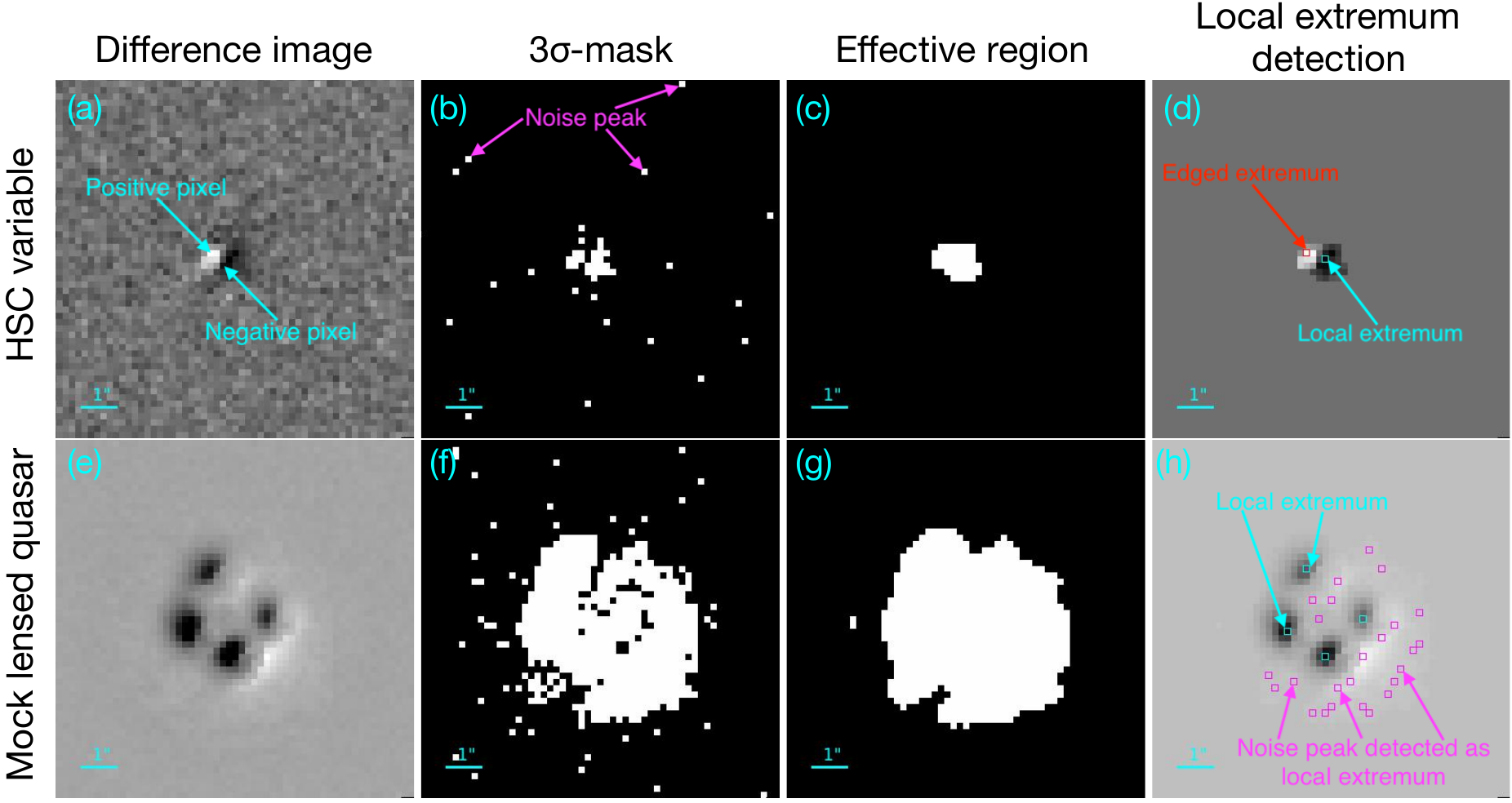}
\caption{Examples of a HSC variable (top row) and an injected lensed quasar (bottom row). From left to right: Difference image, the corresponding 3$\sigma$ mask, the corresponding effective region, and the corresponding local extremum detection. In 3$\sigma$ masks, the white pixels represent the pixels with values larger than 3$\sigma$ or smaller than $-3$$\sigma$ in the difference images. The effective regions denoted by the white pixels show the spatial extent of the objects, after removing noise peaks in the 3$\sigma$ masks. In the local extremum detection, pixels with values larger than all the neighboring pixels or smaller than all the neighboring pixels in the difference images within the effective regions are selected; the local extrema, indicated in cyan, correspond to the "blobs" (See Sec.~\ref{sec:extrema_selection} for details) of the HSC variable/mock lensed quasar, the local extrema indicated in magenta are the noise peaks, and the pixel indicated as "edged extremum" in red is a blob that should be picked as a local extremum, but is missed because it is located at the edge of the effective region. The size of each image cutout is $10\arcsec \times 10\arcsec$.}
\label{fig:var_and_mcok}
\end{figure*}

%
We search for lensed quasars with all the 13 epochs. Assuming we have $M$ HSC variables, we denote $A_{\text{eff}, t}^m$ as the area of the effective region for a HSC variable in epoch $t$ (Table~\ref{tab:HSC_epochs}), where $m=1,...,M$. For each epoch $t$, we compute the percentile $p\%$ for the area of the effective region from all the HSC variables, $A_{\text{eff},t}(p)$:
\begin{equation}
\label{eq:percentile}
p\% \text{ of } M \text{ HSC variables with } A_{\text{eff}, t}^m < A_{\text{eff}, t}(p).
\end{equation}
We note that, the computation of $ A_{\text{eff}, t}(p)$ is based only on the HSC variables, independent of the mock lensed quasars. For each HSC variable $m$, we then count the number of epochs with the area of the effective region larger than $A_{\text{eff},t}(p)$, $N_\text{epoch}^m(p)$. For example, if a HSC variable has $A_{\text{eff}, t}^{m'} > A_{\text{eff}, t}(p_\text{thrs})$ at a threshold of percentile $p_\text{thrs}\%$, for $t=t_1, t_2, \text{ and } t_3$, we say $N_\text{epoch}^{m'}(p_\text{thrs})=3$. Given a threshold of percentile $p_\text{thrs}\%$, we set another threshold, $N_\text{thrs}$, and a HSC variable is classified as a candidate for lensed quasar if $N_\text{epoch}^{m'}(p_\text{thrs}) > N_\text{thrs}$.\\

We also perform the calculation in Eqs.~\ref{eq:3sigma_mask}-\ref{eq:region_area} for each injected mock lensed quasar. Fig.~4e shows an example of injected mock lensed quasar in the same epoch as Fig.~4a. Figs.~4f and 4g are respectively the 3$\sigma$ mask and the effective region of the mock lensed quasar in Fig.~4e. As shown in Figs.~4c and 4g, the injected mock lensed quasar has a much larger area of the effective region.\\  

\subsection{Preselection by “number of blobs” }
\label{sec:extrema_selection}
In addition to the extendedness, we could also apply the "blob-feature" as a secondary criterion to improve the search method. The image residuals of variable sources in the difference image either have positive pixel values or negative pixel values, depending on the brightness change, and the pixel values elsewhere should be zero, up to the noise level. The residuals with positive pixel values are like "white blobs", and the residuals with negative pixel values are like "black blobs", as shown in Figs.~4a and 4e. In the difference image, these blobs are like local extrema in the zero-background. Owing to the multiple point-like image residuals, lensed quasars tend to have a larger number of local extrema, compared to unlensed variable sources. Therefore, we enhance the lens search method in Sec.~\ref{sec:extendedness_only} with a preselection of HSC variables based on a criterion on the number of local extrema.\\ 

We first define a local extremum for the difference image. A pixel $(i, j)$ is a local extremum, if 
\begin{equation}
\label{eq:extrm_begin}
I_\text{mask}(i, j)=1
\end{equation}
and
\begin{equation}
\label{eq:eff_eq}
I_\text{eff}(i, j)=1,
\end{equation}
with its neighboring pixels $(i', j')$ satisfying
\begin{equation}
I_\text{eff}(i', j')=1
\end{equation}
and
\begin{equation}
\label{eq:extrm_end}
I(i,j) > I(i', j') \text{ or } I(i,j) < I(i',j'),
\end{equation}
for $i'=i-1,i,i+1$ and $j'=j-1,j,j+1$, except $(i', j')=(i, j)$. The local extrema are indicated in cyan in Figs.~4d and 4h. Figs.~4d and 4h only keep the pixel values for the pixels satisfying Eq.~\ref{eq:eff_eq} and filter out the other pixels. We note that, with the definition in Eqs.~\ref{eq:extrm_begin}-\ref{eq:extrm_end}, the local extrema do not completely correspond to the blobs. As shown in Fig.~4d, we lose the white blob in the left as a local extremum because the pixel that is supposed to be the local extremum, indicated as "edged extremum" in red, is located at the edge of the region that is effectively related to the HSC variable and thus fails the condition in Eq.~15. Furthermore, the local extrema could also come from noise peaks, as indicated in magenta in Fig.~4h. Even though the local extrema and the blobs are not in one-to-one correspondence, their numbers are still highly correlated.\\     

Before we compute the percentile $A_{\text{eff},t}(p)$ at $p\%$ in Eq.~\ref{eq:percentile}, we count the number of local extrema, $N_\text{extrm}^m$, for each HSC variable $m$ (where $m=1,...,M$) in an epoch of choice. Depending on the imaging survey, this epoch could be for example the best-seeing epoch, or median-seeing epoch.  In the HSC transient survey, the best-seeing epoch with a remarkable value of 0.42 arcsec is so good that the sharp images lead to significant artifacts in the deep difference image that affects the number of extrema. We therefore use the median-seeing epoch (with seeing of 0.72 arcsec) in the HSC transient survey for detecting the local extrema because this is the regime in which the difference imaging pipeline runs well. Given a criterion, $N_\text{crit}$, a HSC variable is discarded from the lens classification if $N_\text{extrm}^{m'} \leq N_\text{crit}$, and a HSC variable is kept in the lens classification if $N_\text{extrm}^{m'} > N_\text{crit}$.\\ 

\begin{figure*}
\centering
\includegraphics[scale=0.6]{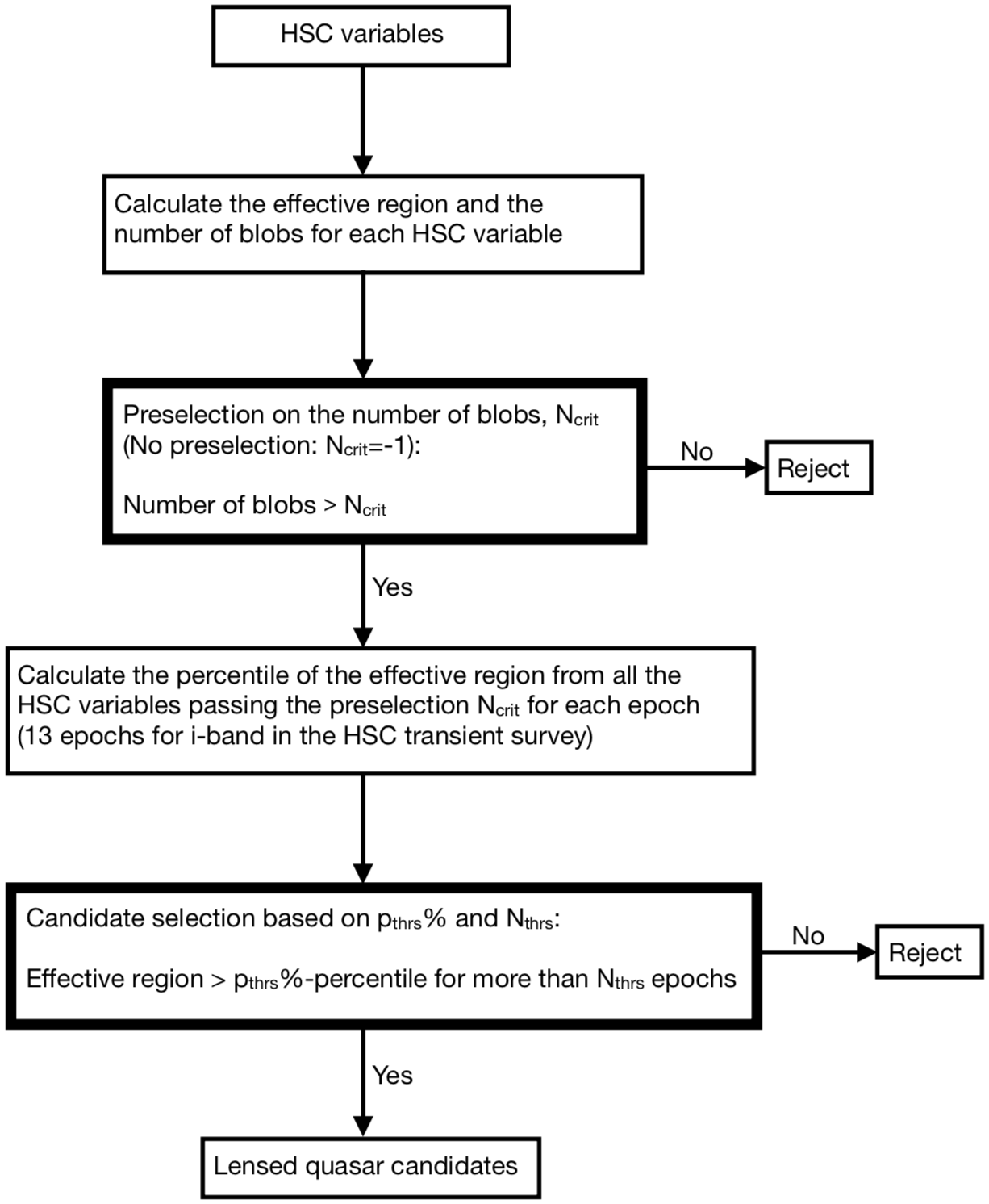}
\caption{Flow chart illustrating the lensed quasar candidate selection procedure. The effective region (Eqs.~\ref{eq:3sigma_mask}-\ref{eq:region_area}) and number of blobs (Eqs.~\ref{eq:extrm_begin}-\ref{eq:extrm_end}) are calculated for each variable object from the HSC transient survey (HSC variable). The HSC variables with the number of blobs larger than $N_\text{crit}$ are preselected. The percentile of the effective region for each epoch from all the HSC variables that remain after the preselection on the number of blobs is calculated. Finally, two thresholds, $p_\text{thrs}\%$ and $N_\text{thrs}$, are set; the HSC variables are selected that have an effective region larger than $p_\text{thrs}\%$ percentile for more than $N_\text{thrs}$ epochs, as lensed quasar candidates.} 
\label{fig:selection_overview}
\end{figure*}

Assuming we have $M'$ HSC variables with their number of the local extrema larger than $N_\text{crit}$ after the preselection (where $M' \leq M$), we compute their area of the effective region for each epoch $t$ ($A_{\text{eff},t}^{m'}$, $m'=1,...,M'$), and further calculate the percentile $p\%$ for the area of the effective region from these $M'$ HSC variables, $A'_{\text{eff}, t}(p)$, in each epoch $t$, which is similar to Eq.~\ref{eq:percentile}. We notice that, at a same percentile $p\%$, $A'_{\text{eff},t}(p)$ is generally larger than $A_{\text{eff},t}(p)$, that is, the percentile $p\%$ from all the HSC variables before the preselection. This is because the HSC variables with fewer local extrema are also the HSC variables with fewer spatially extended blobs counting for the extendedness. Thus when we reject the HSC variables with a small number of the local extrema, we also reject the HSC variables that are small in the area of the effective region. Raising $A_{\text{eff},t}(p)$, the percentile $p\%$ for the area of the effective region, with the preselection by the number of blobs makes the further lens classification more efficient and helps us to avoid large number of false candidates for lensed quasar, although we would lose a few candidates, particularly narrowly separated lensed quasars (Sec.~\ref{sec:performance_extrema}). We illustrate our lensed quasar candidate selection procedure (Sec.~\ref{sec:extendedness_only} and Sec.~\ref{sec:extrema_selection}) in Fig.~\ref{fig:selection_overview}.\\    

\section{Performance test}
\label{sec:performance}
In this section, we examine the performance of our lens search algorithm. To test the performance of the lens search algorithm, we need not only the deep difference images of lensed quasars, but also the deep difference images of the non-lensed objects. We exploit 12910 HSC variables from the HSC transient survey as the non-lensed objects to test our lens search algorithm. Based on the estimation of the lensed quasar population in OM10, we expect that at most 1-2 quad(s) lie in the HSC transient survey according to the survey depth and the covering area. The 12910 HSC variables we use to test our lens search algorithm come from a $\sim$0.64 $\deg^{2}$ sub-sky area in the HSC transient survey (1.77 $\deg^{2}$), therefore all of these HSC variables are likely to be false positives. Like the 2033 injected mock lensed quasars, we take the deep difference images from all the 13 epochs for each HSC variable.\\ 

\subsection{Classification based only on spatial extent}
\label{sec:performance_basic}
The lens search algorithm performance is measured by the true-positive rate (TPR) and the false-positive rate (FPR), which are defined as
\begin{equation}
\text{TPR} = \frac{N_\text{TP}}{N_\text{P}}
\end{equation}
and
\begin{equation}
\text{FPR} = \frac{N_\text{FP}}{N_\text{N}},
\end{equation}
where $N_\text{TP}$ is the number of correctly identified positive cases, $N_\text{P}$ is the number of total positive cases, $N_\text{FP}$ is the number of falsely identified negative cases, and $N_\text{N}$ is the number of total negative cases. In this case, the positive cases are the quasar lenses, and the negative cases are the non-lensed objects, which are the HSC variables from the HSC transient survey.\\

We split the injected mock lensed quasars into four subgroups according to their brightness and quasar-image separation, and test the performance of the lens search algorithm individually for each subgroup. In this work, a lensed quasar is bright if the magnitude of the third brightest image $m_\text{3rd} < 22.0 \text{ mag}$, otherwise this lensed quasar is faint ($22.0 \text{ mag} \leq m_\text{3rd} <  24.0 \text{ mag}$); a lensed quasar has wide separation if the largest separation among the pairs of the lensed images $\theta_\text{LP} > 1.5\arcsec$, otherwise this lensed quasar has narrow separation ($0.5\arcsec < \theta_\text{LP} \leq 1.5\arcsec$). Therefore, the four subgroups are bright lensed quasars with wide separation, bright lensed quasars with narrow separation, faint lensed quasars with wide separation, and faint lensed quasars with narrow separation.\\

For each subgroup, $N_\text{P}$ is the total number of lensed quasars in the subgroup, and $N_\text{TP}$ is the number of lensed quasars in the subgroup that are classified as lens candidates by the lens search algorithm. For example, in the "bright-wide" group, $N_\text{P}$ is the total number of bright lensed quasars with wide separation, and $N_\text{TP}$ is the number of the bright lensed quasars with wide separation that are classified as lens candidates by the lens search algorithm. For all the four subgroups, $N_\text{N}$ is the total number of the HSC variables that we use to test the lens search algorithm ($N_\text{N}=12910$ in this paper), and $N_\text{FP}$ is the number of the HSC variables that are falsely classified as the lens candidates by the lens search algorithm. The quantities $N_\text{N}$ and $N_\text{FP}$ only comprise the HSC variables and they do not correspond to any specific subgroup.\\ 

\begin{figure}
\centering
\includegraphics[width=\columnwidth]{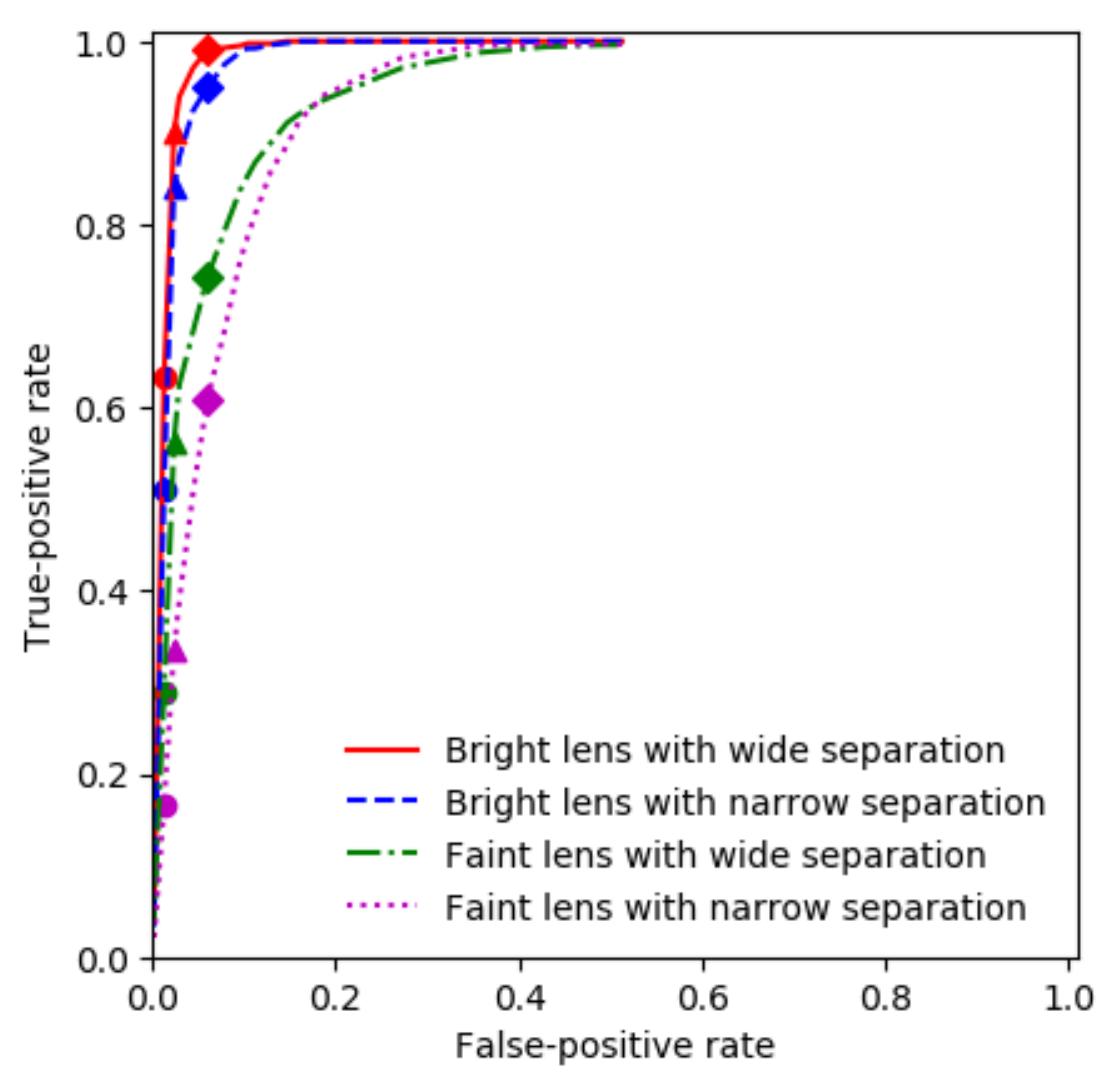}
\caption{Lens search method ROC curves based only on spatial extent. A lens is considered to be bright if its third brightest image has $m_\text{3rd} < 22.0 \text{ mag}$, and faint if $22.0 \text{ mag} \leq m_\text{3rd} < 24.0 \text{ mag}$; a lens is considered to have wide separation if the largest separation among the pairs $\theta_\text{LP} > 1.5\arcsec$, and narrow separation if $0.5\arcsec < \theta_\text{LP} \leq 1.5\arcsec$. The ROC curves are plotted by varying $p_\text{thrs}\%$ from 5\% to 99.5\% at $N_\text{thrs}=9$ (see Sec.~\ref{sec:performance_basic} for details). The diamonds, triangles, and circles indicate the points on the ROC curves when $p_\text{thrs}\%= 90\%, 95\%, \text{ and } 97.5\%$, respectively. The lens search algorithm in this work could detect the bright lensed quasars with wide separation with $(\text{TPR}, \text{FPR})=(90.1\%, 2.3\%)$. The lens search performance for the bright lensed quasars with narrow separation is similar to the bright lensed quasars with wide separation.}
\label{fig:main_without_extrema}
\end{figure}

Fig.~\ref{fig:main_without_extrema} shows the receiver operating characteristic (ROC) curves, TPR against FPR, for the four subgroups. The ROC curves quantitatively show the lens search performance, and the best lens search performance would give points in the top left corner, indicating high TPRs and low FPRs. We can further quantify the lens search performance using the minimum distance between the top left corner ($(\text{TPR}, \text{FPR})=(100\%, 0\%)$) and the ROC curve, $d$. The smaller the value of $d$, the better the performance of the lens search algorithm is. The result in Fig.~\ref{fig:main_without_extrema} is from the method in Sec.~\ref{sec:extendedness_only}, by varying $p_\text{thrs}\%$ at $N_\text{thrs}=9$.\footnote{We tested our lens search algorithm for $N_\text{thrs}=0,1,2,...,12$, and we have the optimal ROC curve when $N_\text{thrs}=9$ (i.e., smallest value of $d$).} Increasing the value of $p_\text{thrs}\%$ from $5\%$ to $99.5\%$ gives points along the ROC curves from the top middle to the bottom left corner. The diamonds, triangles, and circles in Fig.~\ref{fig:main_without_extrema} indicate TPRs and FPRs at $p_\text{thrs}\%= 90\%, 95\%, \text{ and } 97.5\%$, respectively, for each subgroup of the injected mock lensed quasars. We list the values of TPRs, FPRs, and $d$ in Table~\ref{tab:ROC_results_no_ex}, and the numbers of pixels for $A_\text{eff}(p_\text{thrs})$ at $p_\text{thrs}\%= 90\%, 95\%, \text{ and } 97.5\%$ in Table~\ref{tab:percentile_pixels}.\\

\begin{table*}
\centering
\begin{tabular}{lllllllllllll}
\hline
                &  & \multicolumn{2}{l}{Bright-wide} &  & \multicolumn{2}{l}{Bright-narrow} &  & \multicolumn{2}{l}{Faint-wide} &  & \multicolumn{2}{l}{Faint-narrow} \\
\hline                
\multicolumn{1}{c}{$p_\text{thrs}$} &  & TPR            & FPR            &  & TPR             & FPR             &  & TPR            & FPR           &  & TPR             & FPR            \\
\hline
90\%            &  &       0.990         &       0.060       &  &                 0.949 &          0.060    &  &           0.741&               0.060 &  &                 0.608 &            0.060    \\
95\%            &  &       0.901        &        0.023     &  &                   0.840 &            0.023  &  &             0.563 &                0.023 &  &             0.335  &            0.023   \\
97.5\%          &  &      0.632          &      0.012         &  &               0.510  &        0.012        &  &       0.288          &          0.012     &  &         0.165        &        0.012       \\
\hline
\multicolumn{1}{c}{$d$}             &  & \multicolumn{2}{c}{0.053}       &  & \multicolumn{2}{c}{0.078}         &  & \multicolumn{2}{c}{0.170}      &  & \multicolumn{2}{c}{0.182}  \\
\hline
\end{tabular}
\caption{True-positive rates, FPRs, and $d$ (the minimum distance between the ROC curves and the top left corner) of our lens search algorithm at $N_\text{thrs}=9$.}
\label{tab:ROC_results_no_ex}
\end{table*}

\begin{table}[]
\begin{tabular}{c|ccc}
\hline
Observation date     & \multicolumn{3}{c}{$A_\text{eff}(p_\text{thrs})$}\\
\hline                                                                           
\multicolumn{1}{l|}{} & \multicolumn{1}{l}{$p_\text{thrs}=90\%$} & \multicolumn{1}{l}{$p_\text{thrs}=95\%$} & \multicolumn{1}{l}{$p_\text{thrs}=97.5\%$} \\
\hline
2016-11-25           & 78                                 & 119                                & 178                                  \\
2016-11-29           & 94                                 & 163                                & 270                                  \\
2016-12-25           & 140                                & 314                                & 708                                  \\
2017-01-02           & 114                                & 170                                & 306                                  \\
2017-01-23           & 113                                & 163                                & 247                                  \\
2017-01-30           & 84                                 & 124                                & 179                                  \\
2017-02-02           & 105                                & 152                                & 203                                  \\
2017-02-25           & 71                                 & 104                                & 142                                  \\
2017-03-04           & 123                                & 170                                & 241                                  \\
2017-03-23           & 71                                 & 102                                & 142                                  \\
2017-03-30           & 123                                & 194                                & 398                                  \\
2017-04-26           & 136                                & 276                                & 679                                  \\
2017-04-27           & 77                                 & 106                                & 142                                 \\
\hline
\end{tabular}
\caption{Numbers of pixels for $A_\text{eff}(p_\text{thrs})$ at $p_\text{thrs}\%= 90\%, 95\%, \text{ and } 97.5\%$ among the 12910 HSC variables used to test the lens search algorithm.}
\label{tab:percentile_pixels}
\end{table}

We further explore the performance of our lens search algorithm in each subgroup when $N_\text{thrs}=9$. For the bright lensed quasars with wide separation, our lens search algorithm could identify these objects with a TPR $= 90.1\%$ and a FPR $= 2.3\%$, at $p_\text{thrs}\%=95\%$. Although the ROC curve for the bright lensed quasars with narrow separation is slightly lower, our lens search algorithm can still capture these lensed quasars with a TPR $= 84.0\%$ at the same FPR ($=2.3\%$), when $p_\text{thrs}\%=95\%$. This indicates that our lens search algorithm with the method in Sec.~\ref{sec:extendedness_only} is sensitive to the bright lensed quasars regardless of the separation. Given the much lower ROC curves, the faint lensed quasars are mainly more difficult to detect by our lens search algorithm. For the faint lensed quasars with wide separation, the TPR hugely drops to $56.3\%$ at $p_\text{thrs}\%=95\%$, while the nearest point to the top left corner on the ROC curve is $(\text{TPR}, \text{FPR})=(89.0\%, 13.0\%)$ at $p_\text{thrs}\%=82\%$. The false positives that our lens search method can possibly detect are very bright objects, such as very bright galaxies or very bright variable stars. Those objects are so bright that the transient data pipeline might not be able to perform the image subtraction properly, and their difference images might not represent the correct brightness change.\\ 

\begin{figure}
\centering
\includegraphics[width=\columnwidth]{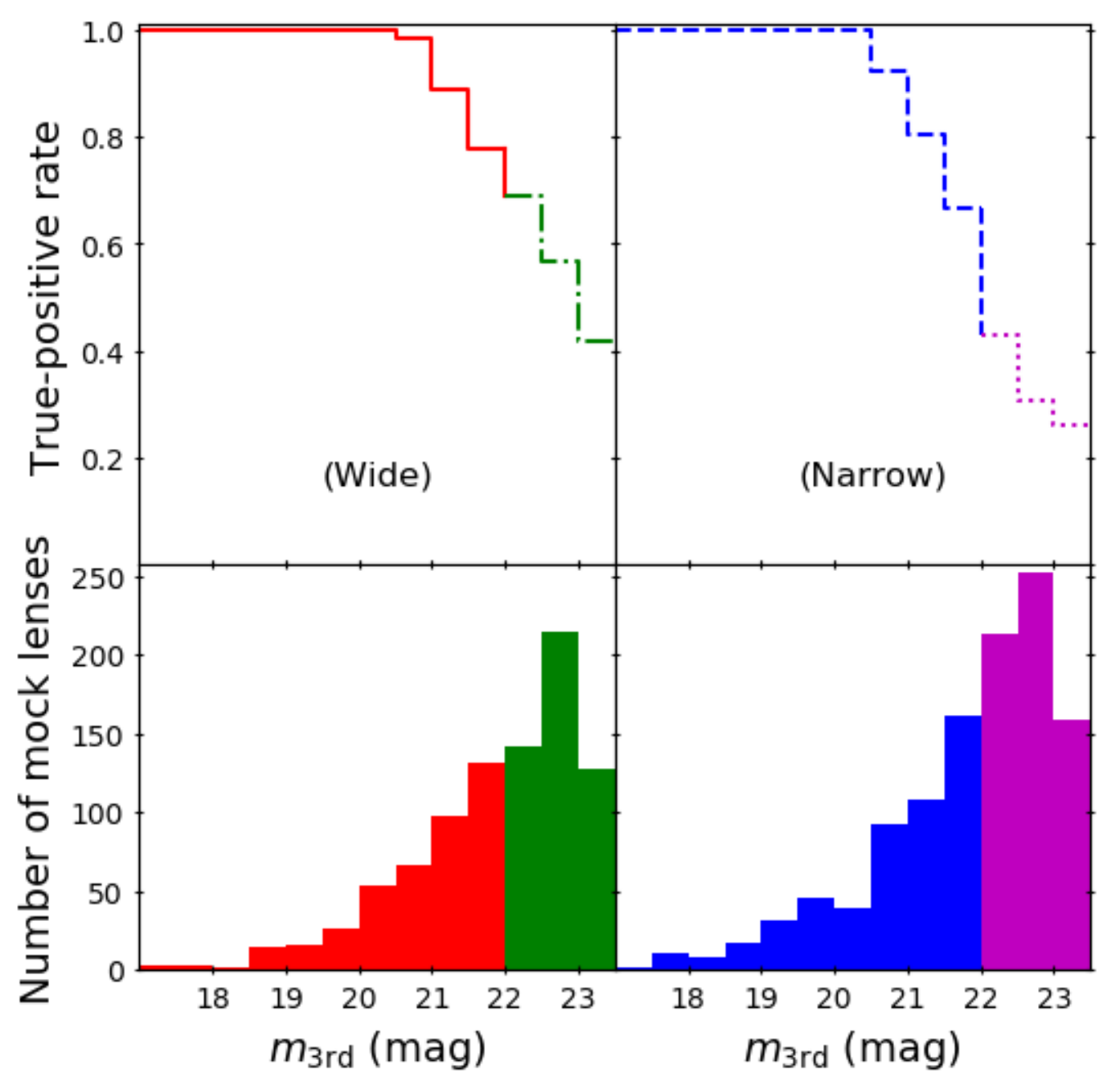}
\caption{Sensitivity to the quasar image brightness $m_\text{3rd}$ of the lens search method based only on spatial extent. Wide: The group of lensed quasars with the largest separation among the pairs $\theta_\text{LP} > 1.5\arcsec$. Narrow: The group of lensed quasars with $0.5\arcsec < \theta_\text{LP} \leq 1.5\arcsec$. The top panels show the TPRs of each $m_\text{3rd}$ bin, when $p_\text{thrs}\%=95\%$ and $N_\text{thrs}=9$. The bottom panels show the number of mock lenses in each $m_\text{3rd}$ bin. Our lens search algorithm can identify more than 90\% of the lensed quasars with $m_\text{3rd} < 21.0 \text{ mag}$.} 
\label{fig:sensitivity_bright}
\end{figure}
We examine our lens search algorithm's sensitivity to $m_\text{3rd}$ with $p_\text{thrs}\%=95\%$ and $N_\text{thrs}=9$ in Fig.~\ref{fig:sensitivity_bright}. The top panels in Fig.~\ref{fig:sensitivity_bright} show the TPRs, and the bottom panels show the number of mock lenses in each $m_\text{3rd}$ bin. Our lens search algorithm can detect all the lensed quasars with the third brightest image brighter than 20.5 mag ($m_\text{3rd}< 20.5 \text{ mag}$), and the TPRs for the lensed quasars with $m_\text{3rd} < 21.0 \text{ mag}$ are larger than or equal to $90\%$. We show again that our lens search algorithm with the method from Sec.~\ref{sec:extendedness_only} has similar sensitivity to the bright lensed quasars with both wide and narrow separation; this is because most of the HSC variables are unlensed sources and have small areas of the effective region, resulting in small values for the percentile of $p_\text{thrs}\%$. Therefore, a bright lensed quasar typically has an area of the effective region larger than the percentile of $p_\text{thrs}\%$ and can be identified as a lens candidate regardless of its separation. The TPRs gradually drop when $m_\text{3rd} \geq 21.0 \text{ mag}$. Our lens search algorithm has poorer performance on the faint lensed quasars because the difference images of the faint lensed quasars have fewer pixels with values larger than $3\sigma$ (Eq.~\ref{eq:3sigma_mask}). Because the brightness change of quasar is generally small, the area of the effective region in each epoch, $A_{\text{eff},t}^m$ (Eqs.~\ref{eq:region_area} and \ref{eq:percentile}, see Sec.~\ref{sec:extendedness_only} for more detail), is also small even for the lenses with wide separation, which makes the faint lensed quasars unable to pass the given thresholds, $p_\text{thrs}\%$ and $N_\text{thrs}$.\\

\subsection{Classification using both spatial extent and number of blobs}
\label{sec:performance_extrema}
\begin{figure}
\centering
\includegraphics[width=\columnwidth]{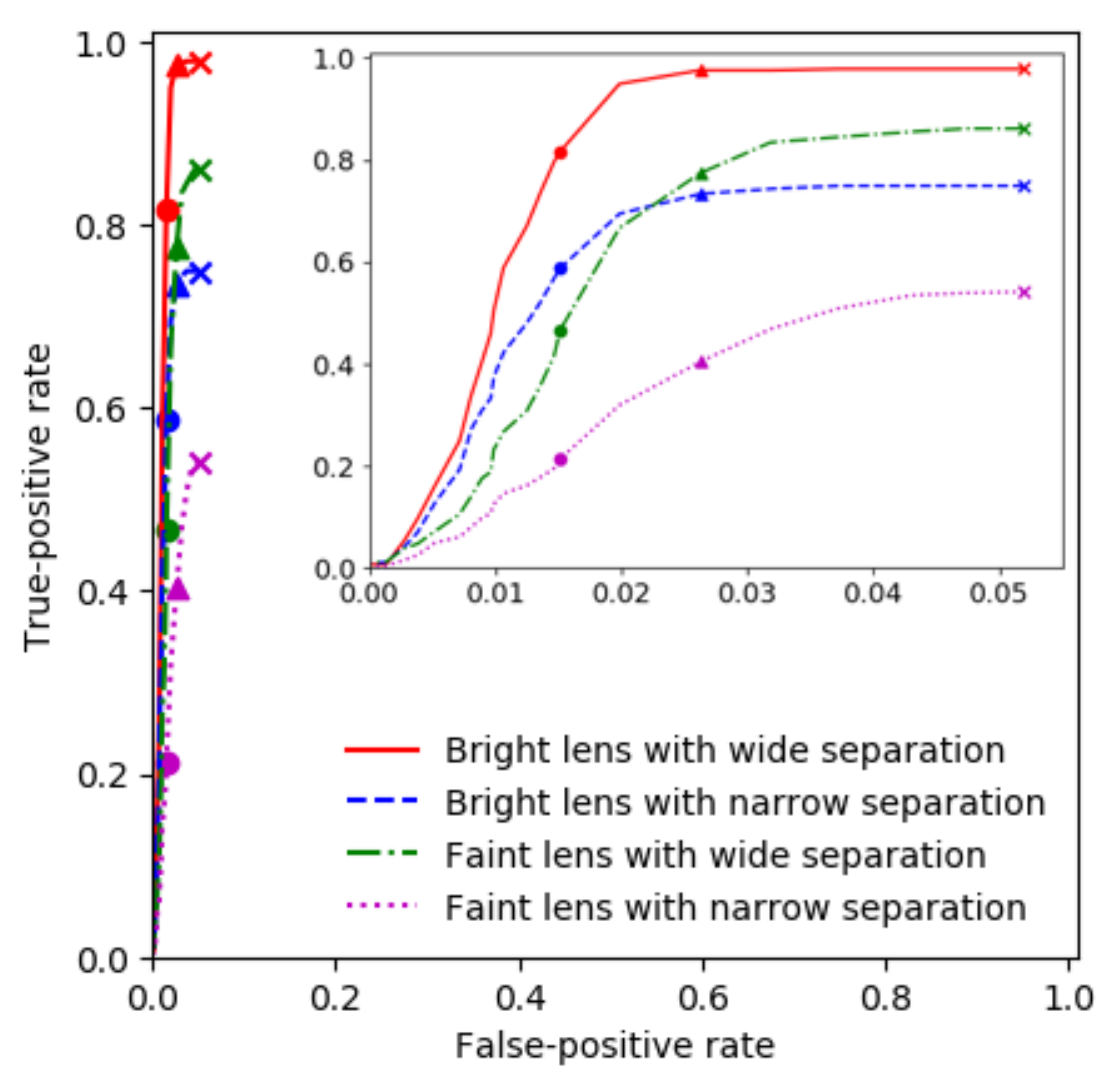}
\caption{ROC curves of the lens search method using both spatial extent and number of blobs. The subgroups of lensed quasars are the same as the subgroups in Fig.~\ref{fig:main_without_extrema}. The ROC curves are plotted by varying $p_\text{thrs}\%$ from 0\% to 99.5\% at $N_\text{thrs}=4$ with the preselection on the number of blobs $N_\text{crit}=2$. The crosses, triangles, and circles indicate the points on the ROC curves when $p_\text{thrs}\%=0\%, 55\%, \text{ and } 75\%$, respectively. With the preselection $N_\text{crit}=2$, our lens search algorithm could identify the bright lensed quasars with wide separation at $(\text{TPR}, \text{FPR}) = (97.6\%, 2.6\%)$, which is similar to the lens search performance when no preselection is performed. The ROC curves also start from a much lower FPR (crosses) with the preselection $N_\text{crit}=2$. The ROC curves are zoomed-in with the FPR ($x$-axis) spanning from 0 to 0.05 in the small panel.}
\label{fig:main_with_extrema}
\end{figure}

Now we explore the lens search performance with the preselection based on the number of blobs. We test how the ROC curves change with the criterion number of local extrema, $N_\text{crit}$ (see Sec.~\ref{sec:extrema_selection} for details). In Fig.~\ref{fig:main_with_extrema}, we plot the ROC curves by varying $p_\text{thrs}\%$ for $N_\text{crit}=2$. As shown in Fig.~\ref{fig:main_with_extrema}, after the preselection is applied ($N_\text{crit}=2$), our lens search algorithm can capture the bright lensed quasars with wide separation at $(\text{TPR}, \text{FPR}) = (97.6\%, 2.6\%)$ with thresholds, $p_\text{thrs}\%=55\%$ and $N_\text{thrs}=4$.\footnote{When the preselection $N_\text{crit}=2$ is applied, we have the optimization of the ROC curves at $N_\text{thrs}=4$.} Comparing to the lens search performance without the preselection by the number of blobs ($N_\text{crit}=0$), the lens search algorithm applying the preselection ($N_\text{crit} > 0$) gives a similar performance with much looser constraints on $p_\text{thrs}\%$ and $N_\text{thrs}$. The crosses, triangles, and circles in Fig.~\ref{fig:main_with_extrema} represent TPRs and FPRs at $p_\text{thrs}\%= 0\%, 55\%, \text{ and } 75\%$, respectively. With the preselection by number of blobs, the ROC curves start from a much lower FPR, compared to the ROC curves without the preselection ($N_\text{crit}=0$). The crosses in Fig.~\ref{fig:main_with_extrema} indicate that when $N_\text{crit}=2$, we have a much lower FPR ($\sim$5\%) at $N_\text{thrs}=4$, even when there is no constraint from $p_\text{thrs}\%$. We list the values of TPRs, FPRs, and $d$ (the minimum distance between the ROC curves and the top left corner) from the lens search performance with the preselection by the number of blobs in Table~\ref{tab:ROC_results_w_ex}. The numbers of pixels for $A_\text{eff}(p_\text{thrs})$ at $p_\text{thrs}\%=55\% \text{ and }75\%$ after the preselection are listed in Table~\ref{tab:percentile_pixels_w_ex}.\\ 

We further look into each subgroup of lensed quasars to compare the performance between the two conditions, with and without the preselection by number of blobs ($N_\text{crit} > 0$ and $N_\text{crit} = 0$). At small values of $p_\text{thrs}\%$, the TPRs are $\sim$100\% for all the four subgroups when $N_\text{crit} = 0$, while the TPRs for the four subgroups all drop when $N_\text{crit} = 2$. This indicates that, although the preselection based on number of blobs hugely decrease the FPRs, the preselection also discards a part of the candidates for lensed quasars. The TPR decreases by $\sim$2.2\% for the bright lensed quasars with wide separation, by $\sim$25.1\% for the bright lensed quasars with narrow separation and by $\sim$13.9\% for the faint lensed quasars with wide separation. We notice that, unlike Fig.~\ref{fig:main_without_extrema} in which the ROC curve of the bright lensed quasars with narrow separation is higher than the ROC curve of the faint lensed quasars with wide separation, when $N_\text{crit} = 2$, the ROC curve of the faint lensed quasars with wide separation becomes higher than the ROC curve of the bright lensed quasars with narrow separation (at FPR $\gtrsim$0.02). This implies the blob-based preselection has a more significant impact on the lensed quasars with narrow separation.\\

\begin{figure}
\centering
\includegraphics[width=\columnwidth]{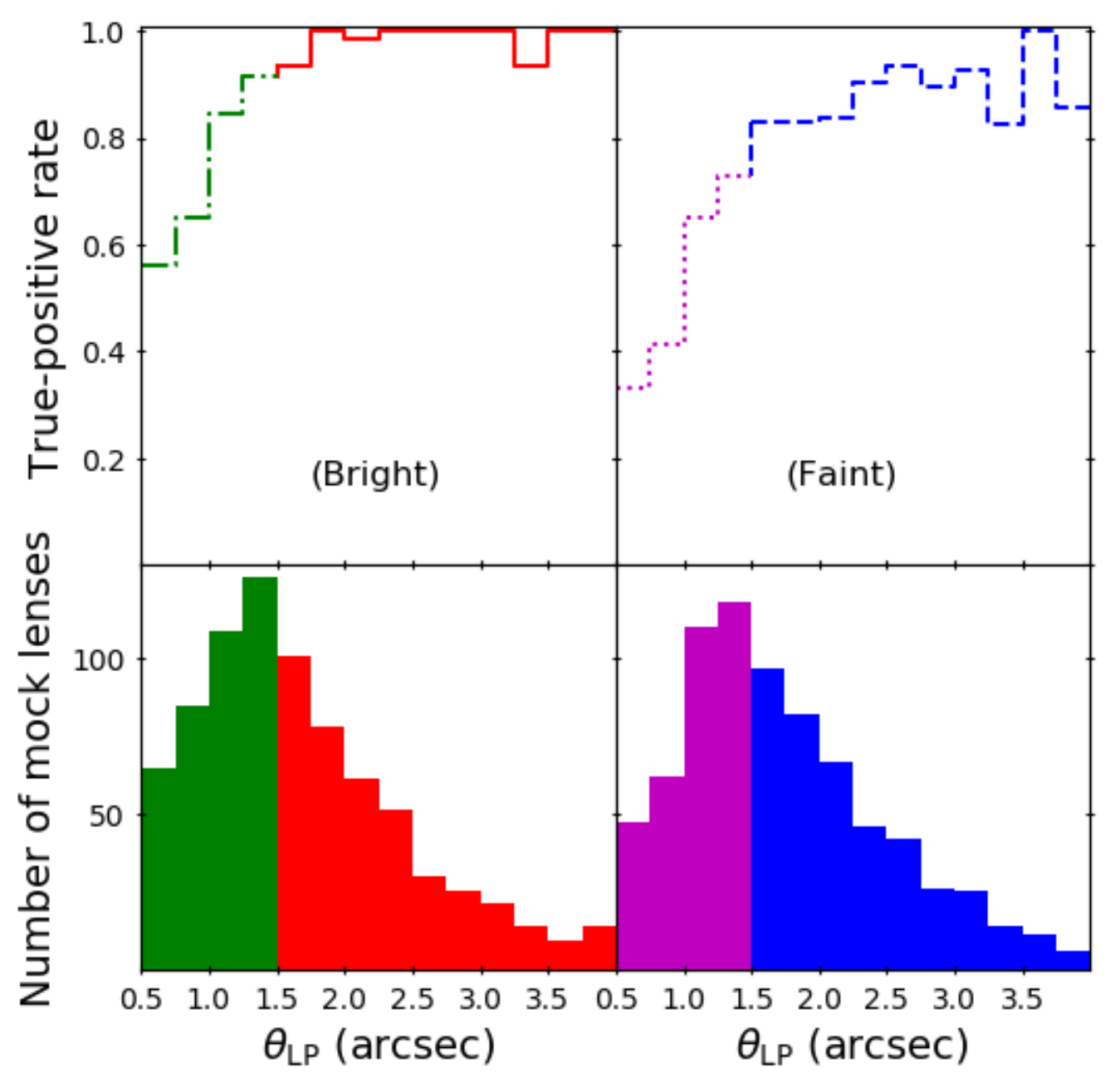}
\caption{Sensitivity to the quasar image separation $\theta_\text{LP}$ of the lens search method using both spatial extent and the number of blobs. Bright: The group of lensed quasars with the third brightest image brightness $m_\text{3rd} < 22.0 \text{ mag}$. Faint: The group of lensed quasars with $22.0 \text{ mag} \leq m_\text{3rd} < 24.0 \text{ mag}$. The top panels show the TPRs of each $\theta_\text{LP}$ bin, when $p_\text{thrs}\% = 0\%$ and $N_\text{thrs} = 4$, with the preselection by the number of blobs $N_\text{crit} = 2$. The bottom panels show the number of mock lenses in each $\theta_\text{LP}$ bin.}
\label{fig:sensitivity_sep}
\end{figure}

\begin{figure}
\centering
\includegraphics[width=\columnwidth]{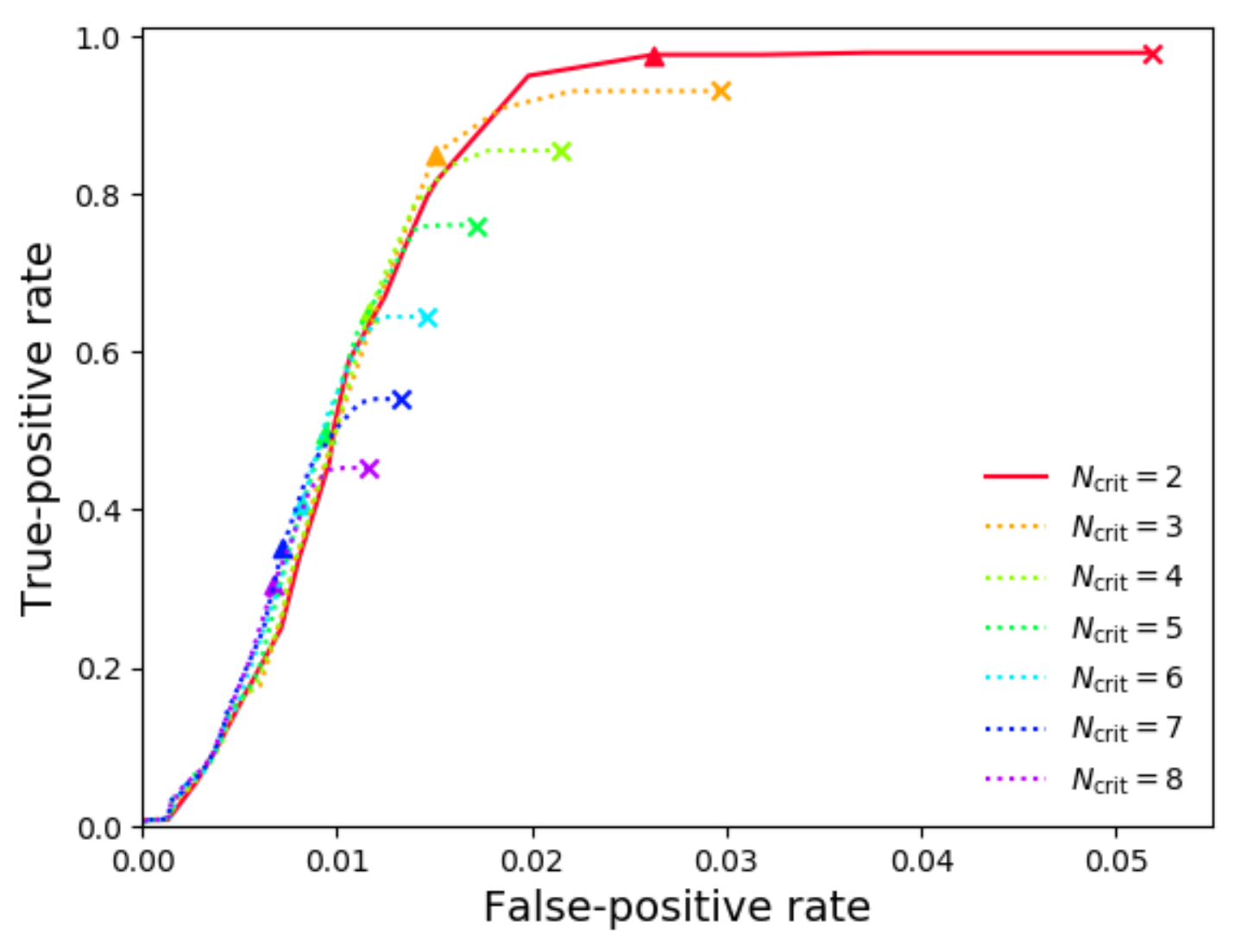}
\caption{Bright lensed quasars ($m_\text{3rd} < 22.0 \text{ mag}$) ROC curves with wide separation ($\theta_\text{LP} > 1.5\arcsec$) of the lens search method with the preselection on the number of blobs at different $N_\text{crit}$ values.}
\label{fig:bright_wide_multi_extrema}
\end{figure}

To investigate the influence of the preselection, we examine the sensitivity to $\theta_\text{LP}$ of the lens search algorithm setting $p_\text{thrs}\%=0\%$, at $N_\text{thrs}=4$, with a preselection $N_\text{crit}=2$. By setting $p_\text{thrs}\%=0\%$, we could study the impact from the preselection more clearly. Fig.~\ref{fig:sensitivity_sep} shows that, when there is no constraint from $p_\text{thrs}$, our lens search algorithm with the preselection by number of blobs could detect the lensed quasars with $\theta_\text{LP} > 1.5\arcsec$ at relatively stable TPRs ($> 80\%$), and there is a great drop in TPRs when $\theta_\text{LP}$ becomes smaller than $1.0\arcsec$. The lensed quasars with narrow separation become even harder to capture when the preselection is applied and this is because, as lensed images with narrow separation are blended together, the number of blobs decreases, making the lensed quasars with narrow separation difficult to meet the given criterion, $N_\text{crit}$. We show the ROC curves for the bright lensed quasars with wide separation at different values of $N_\text{crit}$ (when $N_\text{thrs}=4$) in Fig.~\ref{fig:bright_wide_multi_extrema}, zoomed-in with the $x$-axis spanning from 0 to 0.05. The crosses and triangles in Fig.~\ref{fig:bright_wide_multi_extrema} represent $p_\text{thrs}\% = 0\% \text{ and } 55\%$, respectively. As $N_\text{crit}$ increases, the ROC curve drops and slightly shifts to the left. Both the TPRs and the FPRs decrease when we raise $N_\text{crit}$. 
However, FPR decreases faster than TPR at low values of $N_\text{crit}$.  In particular, we can decrease the FPR by a factor $\sim$2 and have a more efficient lens search when we raise $N_\text{crit}$ from 2 to 3, at the expense of lowering TPR by only $\sim$$5\%$.  Raising $N_\text{crit}$ beyond 3 starts to lead to more decrease in TPR compared to FPR, and is not as advantageous.

\begin{table*}
\centering
\begin{tabular}{lllllllllllll}
\hline
                &  & \multicolumn{2}{l}{Bright-wide} &  & \multicolumn{2}{l}{Bright-narrow} &  & \multicolumn{2}{l}{Faint-wide} &  & \multicolumn{2}{l}{Faint-narrow} \\
\hline                
$p_\text{thrs}\%$ &  & TPR            & FPR            &  & TPR             & FPR             &  & TPR            & FPR           &  & TPR             & FPR            \\
\hline
0\%            &  &       0.978         &       0.052       &  &                 0.749 &          0.052    &  &           0.861&               0.052 &  &                 0.541 &            0.052   \\
55\%            &  &       0.976        &        0.026     &  &                   0.733 &            0.026  &  &             0.774 &                0.026 &  &             0.404  &            0.026   \\
75\%          &  &      0.816          &      0.015         &  &               0.588  &        0.015        &  &       0.466          &          0.015     &  &         0.213        &        0.015       \\
\hline
\multicolumn{1}{c}{$d$}             &  & \multicolumn{2}{c}{0.036}       &  & \multicolumn{2}{c}{0.254}         &  & \multicolumn{2}{c}{0.147}      &  & \multicolumn{2}{c}{0.462}  \\
\hline
\end{tabular}
\caption{Values of TPRs, FPRs, and $d$ (the minimum distance between the ROC curves and the top left corner) at $N_\text{thrs}=4$, when the preselection based on the number of blobs, $N_\text{crit}=2$, is applied.}
\label{tab:ROC_results_w_ex}
\end{table*}

\begin{table}[]
\centering
\begin{tabular}{c|cc}
\hline
Observation date      & \multicolumn{2}{c}{$A_\text{eff}(p_\text{thrs})$}  \\ 
\hline                               
\multicolumn{1}{l|}{} & \multicolumn{1}{l}{$p_\text{thrs}=55\%$} & \multicolumn{1}{l}{$p_\text{thrs}=75\%$} \\
\hline
2016-11-25            & 113                                & 223                                \\
2016-11-29            & 132                                & 292                                \\
2016-12-25            & 167                                & 553                                \\
2017-01-02            & 154                                & 303                                \\
2017-01-23            & 160                                & 283                                \\
2017-01-30            & 120                                & 208                                \\
2017-02-02            & 127                                & 184                                \\
2017-02-25            & 124                                & 208                                \\
2017-03-04            & 157                                & 282                                \\
2017-03-23            & 97                                 & 153                                \\
2017-03-30            & 156                                & 400                                \\
2017-04-26            & 151                                & 568                                \\
2017-04-27            & 99                                 & 161                               \\
\hline
\end{tabular}
\caption{Numbers of pixels for $A_\text{eff}(p_\text{thrs})$ at $p_\text{thrs}\%= 55\%, \text{ and } 75\%$ after applying the preselection on the number of blobs, $N_\text{crit}=2$.}
\label{tab:percentile_pixels_w_ex}
\end{table}

%

\subsection{Classification based on one single epoch}
\label{sec:single_epo}
It is ideal and most efficient to detect lensed quasars through only one epoch (in addition to the reference image). If the lens search algorithm is able to identify candidates for lensed quasars in a single epoch of an ongoing cadenced survey, possible spectroscopic follow-up could be conducted immediately for the confirmation and the lens model, and a further monitoring observation for measuring the time delays could start right away. In this section, we study the performance of our lens search algorithm to capture lensed quasars in one single epoch.\\

\begin{figure*}
\centering
\begin{subfigure}[b]{0.33\textwidth}
\includegraphics[width=\textwidth]{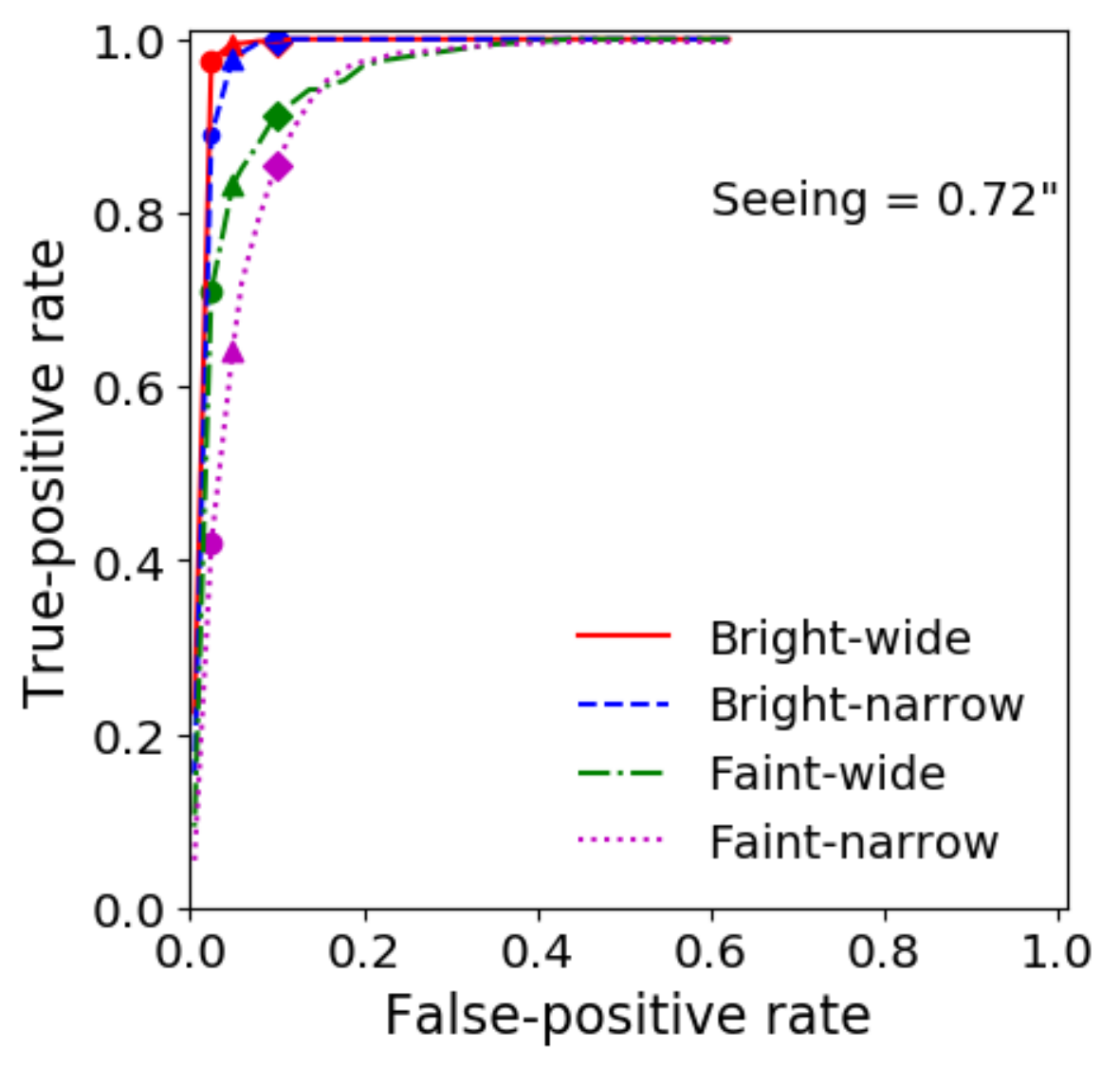}
\end{subfigure}
\begin{subfigure}[b]{0.33\textwidth}
\includegraphics[width=\textwidth]{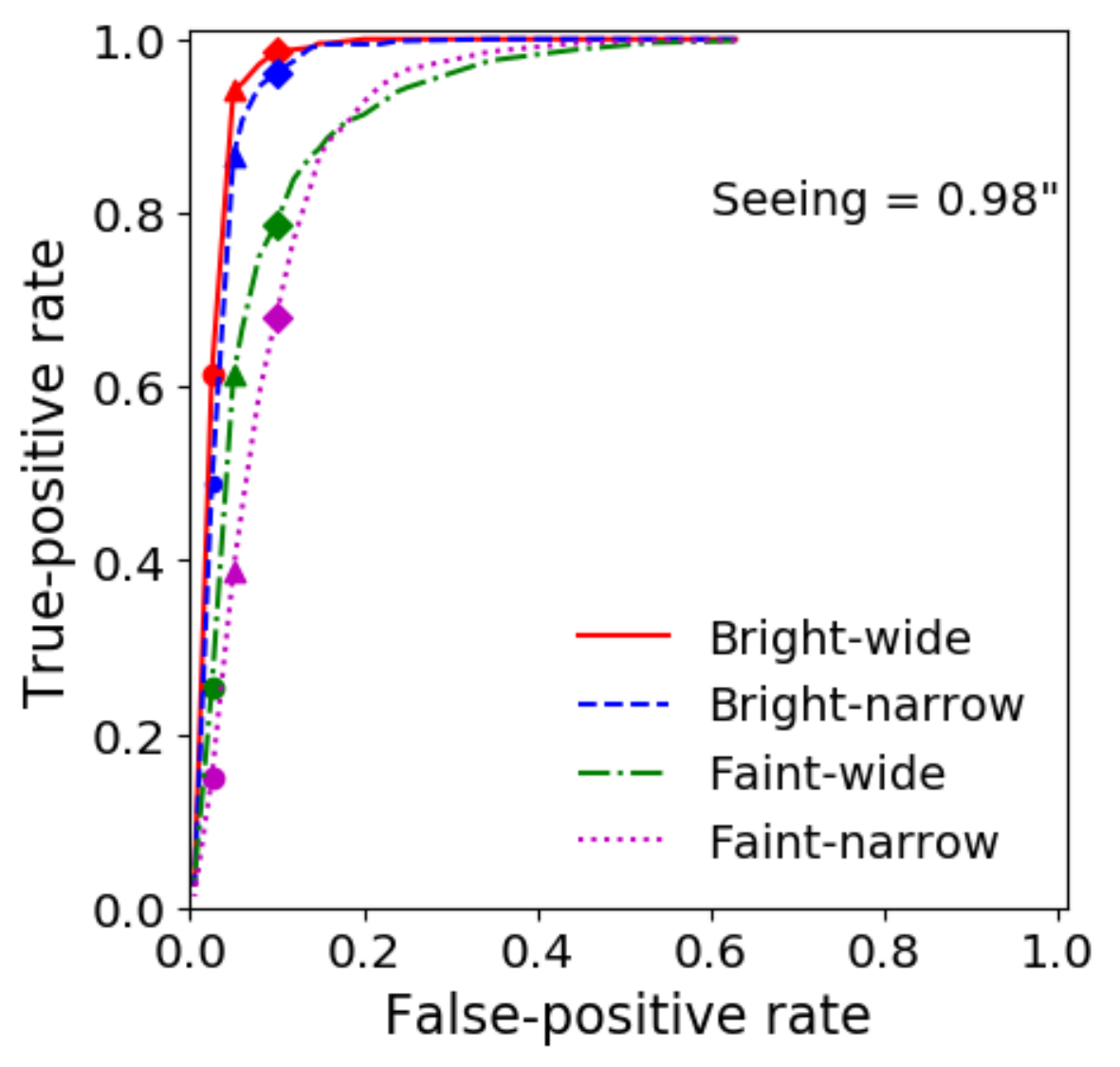}
\end{subfigure}
\begin{subfigure}[b]{0.33\textwidth}
\includegraphics[width=\textwidth]{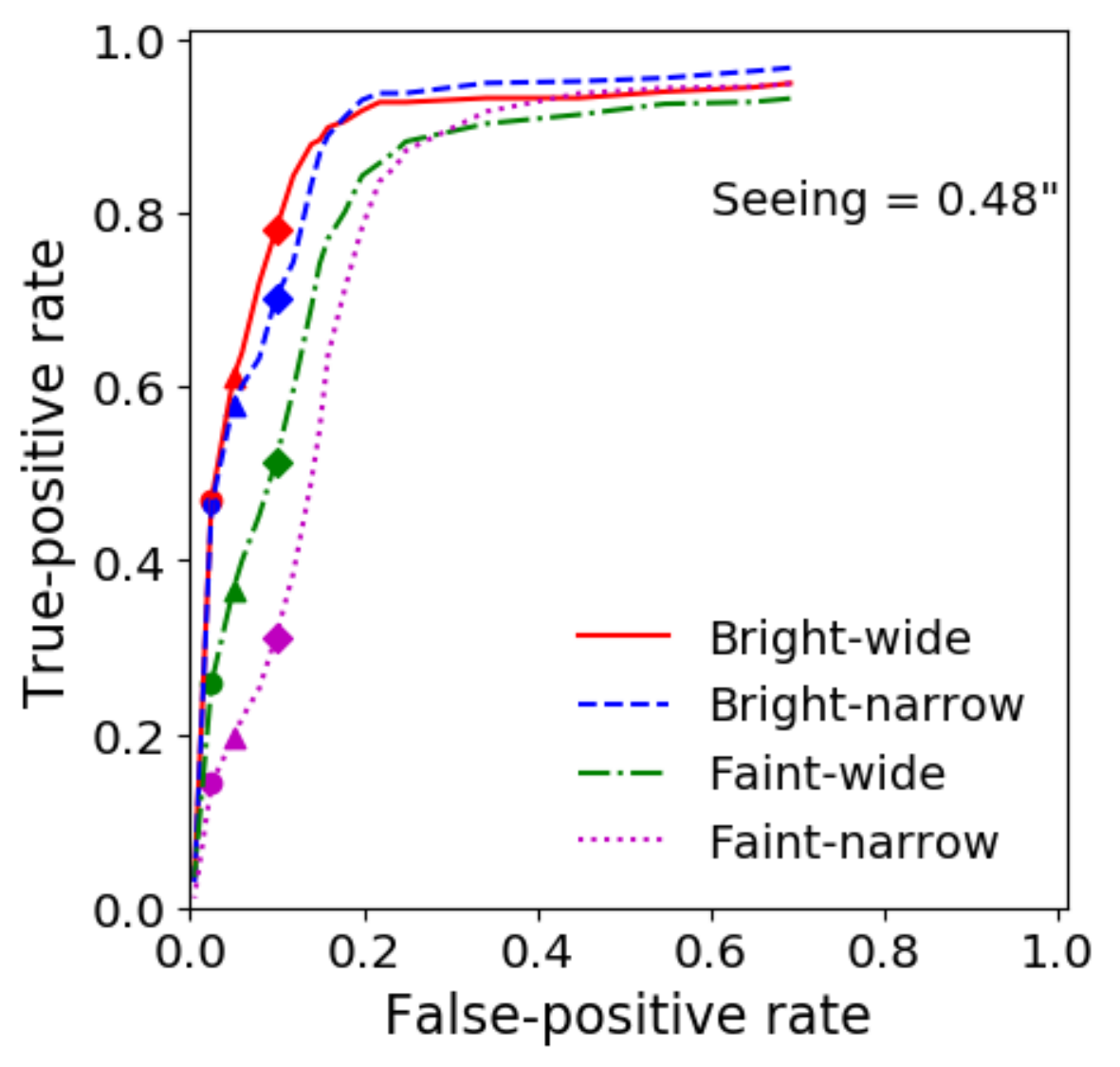}
\end{subfigure}
\caption{Lens search performance based on a single epoch. The lens search algorithm is tested using difference images from only one single epoch for all the 13 epochs in the HSC transient survey; the ROC curves of the epochs with the best performance (left), the average performance (middle), and the worst performance (right) are shown. For lens search algorithm based on a single epoch, only $p_\text{thrs}$ is used as constraint. In each panel, the ROC curves are plotted by varying $p_\text{thrs}\%$; the diamonds, triangles, and circles indicate the points on the ROC curves at $p_\text{thrs}\%=90\%, 95\%, \text{ and } 97.5\%$. The seeing of each epoch is also indicated in each panel. The best lens search performance happened in the single epoch with a seeing of $0.72\arcsec$, capturing the bright lensed quasars with wide separation at $(\text{TPR}, \text{FPR}) = (97.6\%, 2.4\%)$ when $p_\text{thrs}\% = 97.5\%$. In a single epoch with average seeing ($\sim$0.98$\arcsec$), the lens search algorithm can detect the bright lensed quasars with wide separation at $(\text{TPR}, \text{FPR}) = (94.2\%, 5.0\%)$ when $p_\text{thrs}\% = 95.0\%$. The worst lens search performance happens in the epoch with an exceptional seeing value of $0.48\arcsec$ due to artifacts appearing in the difference image pipeline when the seeing is "too" good.} 
\label{fig:performance_sinepo}
\end{figure*}

If only one epoch is available, our lens search algorithm employs only $p_\text{thrs}$. Since the seeing of a single epoch might not be suitable for counting the local extrema, we do not apply the preselection using number of blobs. For one single epoch $t'$, our lens search algorithm classifies a HSC variable $m'$ as a candidate for lensed quasar if 
\begin{equation}
A_{\text{eff},t'}^{m'} > A_{\text{eff},t'}(p_{\text{thrs}}),
\end{equation}  
where $A_{\text{eff},t'}^{m'}$ is the area of the effective region in epoch $t'$ for the HSC variable $m'$, and $A_{\text{eff},t'}(p_{\text{thrs}})$ is the percentile $p_\text{thrs}\%$ for the area of the effective region from all the HSC variables in epoch $t'$. We test the lens search performance in each epoch from the HSC transient survey and show the ROC curves with varied values of $p_\text{thrs}\%$ from three epochs with different seeings in Fig.~\ref{fig:performance_sinepo}. The result in the left panel of Fig.~\ref{fig:performance_sinepo} is from the epoch with the best lens search performance among the 13 epochs, where the nearest point to the top left corner on the ROC curve for the bright lensed quasars with wide separation is $(\text{TPR}, \text{FPR}) = (97.6\%, 2.4\%)$ at $p_\text{thrs}\% = 97.5\%$, which is similar to the performance when all the 13 epochs is used. The middle panel of Fig.~\ref{fig:performance_sinepo} shows the average case of the lens search performance we have in one single epoch, with the nearest point to the top left corner on the ROC curves, $(\text{TPR}, \text{FPR}) = (94.2\%, 5.0\%)$ at $p_\text{thrs}\% = 95.0\%$. The worst performance we have among the 13 epochs is presented in the right panel of Fig.~\ref{fig:performance_sinepo}.\\

\begin{figure}
\centering
\includegraphics[width=\columnwidth]{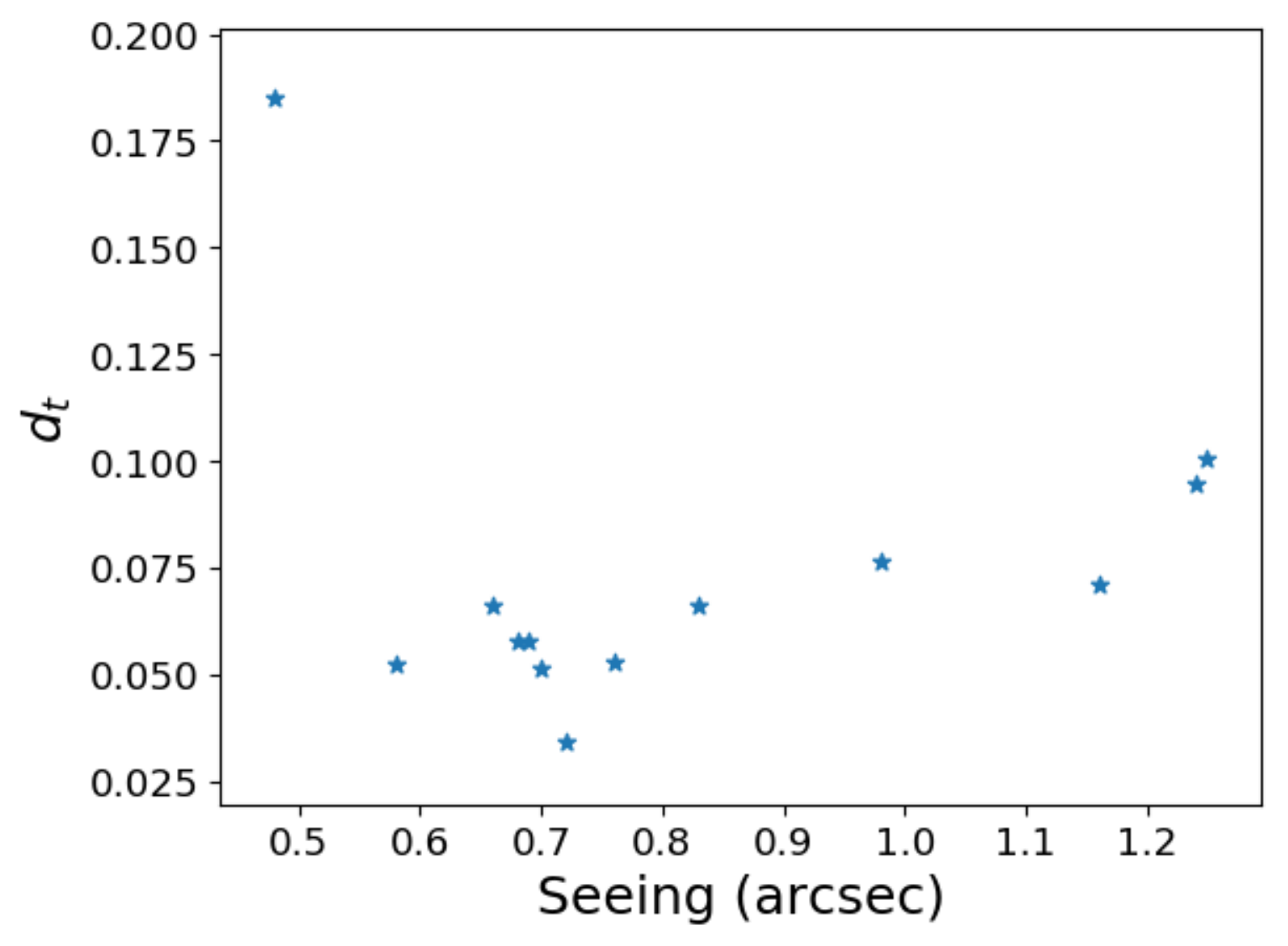}
\caption{Relation between the lens search performance and the seeing in each single epoch from the HSC transient survey. The quantity $d_t$ is defined as the minimum distance between $(\text{TPR}, \text{FPR}) = (100\%, 0\%)$ and the ROC curve of the bright lensed quasars ($m_\text{3rd} < 22.0 \text{ mag}$) with wide separation ($\theta_\text{LP} > 1.5\arcsec$) for each epoch $t$. The 13 star symbols indicate the $d_t$ values for the 13 epochs, plotted against their corresponding seeing. Our lens search algorithm performs well when the single epoch has seeing $\sim$0.7$\arcsec$. The low lens search performance in the epoch with extraordinary seeing ($< 0.5\arcsec$) is due to the significant artifacts in the difference image from the sharp image.}
\label{fig:performance_vs_seeing}
\end{figure}

We further investigate the relation between the lens search performance and the seeing in one single epoch. For each epoch $t$, we define the distance $d_t$ as the distance between $(\text{TPR}, \text{FPR}) = (100\%, 0\%)$ and the nearest point on the ROC curve of the bright lensed quasars with wide separation. The smaller $d_t$, the better the lens search performance is. We plot $d_t$ against the seeing for each epoch $t$ in Fig.~\ref{fig:performance_vs_seeing}. The quantity $d_t$ becomes small when seeing is around $0.7$ arcsec, indicating that our lens search algorithm has good performance when the single epoch has seeing around $0.7$ arcsec. In Fig.~\ref{fig:performance_vs_seeing}, we also see that $d_t$ suddenly increases when seeing is better than $0.5$ arcsec, which verifies that when the seeing is too good, our lens search performance is heavily affected by the significant artifacts from the sharp images.\\    

\section{Conclusions and discussion}
\label{sec:conclusion}
In this work, we present a comprehensive simulation pipeline of time-varied lensed images and develop a new algorithm for searching lensed quasars through their time variability. The simulation pipeline in this work is useful not only for lensed quasar search, but also for other studies, such as studying lensed supernovae and testing lens mass modeling. Our lens search method builds upon the method first proposed by \citet{Kochanek_method}, and provides a practical way of selecting lensed quasar candidates through their difference images.
We summarize the main results as follows:
\begin{enumerate}
\item[--] Our simulation pipeline generates images of lensed quasar accounting for the quasar variability, quasar host, lens galaxy, and the PSF variation. The application of the simulation pipeline to the HSC transient survey yields HSC-like difference images of the mock lensed quasars.
 
\item[--] The lens search algorithm we develop in this work quantifies spatial extent of the variable objects in difference images, and further selects the variable objects with large spatial extent as lensed quasar candidates. The test on a sample containing mock lensed quasars and HSC variables shows that our lens search algorithm could identify the bright lensed quasars with wide separation ($m_\text{3rd} < 22.0 \text{ mag}$ and $\theta_\text{LP} > 1.5\arcsec$) at a high TPR (90.1\%) and a low FPR (2.3\%).   

\item[--] With a preselection on number of blobs, our lens search algorithm could achieve an even higher TPR (97.6\%) without a significant change in FPR (2.6\%) for the bright lensed quasars with wide separation. The preselection is more sensitive to the lensed quasars with wide separation.

\item[--] Although our lens search algorithm mainly uses difference images from multiple epochs, it also works with only one single epoch. The lens search performance in one single epoch depends on the seeing. When the seeing is around 0.7 arcsec, we have the best lens search performance, $(\text{TPR}, \text{FPR}) = (97.6\%, 2.4\%)$ for the bright lensed quasars with wide separation. If the seeing is substantially better or worse than 0.7 arcsec, our lens search performance become poorer.  
\end{enumerate}

Our lens search algorithm will be even more powerful when combined with other lens search techniques. While our lens search algorithm achieves a high TPR and a low FPR, the number of the input variables (e.g., the HSC variables in this work) are typically huge, and thus the absolute number of false-positive lens candidates are also large even with FPR of $1-2\%$, making further confirmation inaccessible. Although the lens search techniques in previous works use static approaches without any information from quasar variability, their exploration in catalogs and image configuration of lensed quasar are helpful for making our lens search algorithm more efficient. For example, we could use the techniques in the cuts of magnitude and color \citep[e.g.][]{Adriano_method, Peter_Williams_method} or \textit{Gaia} multiplets to first select objects that are possible to be lensed quasar and then apply our lens search algorithm to examine their variable nature. Alternatively, after selecting the lensed quasar candidates by our lens search algorithm, we could apply the technique in \citet{chitah} to filter the candidates for lensed quasar. The variability across multiple bands could also be very useful for enhancing our lens search algorithm, while so far our lens search algorithm has exploited only \textit{i} band. For example, we could apply the method based on the spatial extendedness in Sec.~\ref{sec:extendedness_only} to the other bands and select the variable objects appearing as spatially extended in the difference images across multiple bands. Doing so could decrease the number of false-positive lens candidates from one single band. The thresholds on the effective region, $p_\text{thrs}\%$, and the number of epochs, $N_\text{thrs}$, for the other bands depend on the survey, and we need the mock lenses for the other bands to investigate them. We will explore the multiband variability in the future work.\\

Since the lens search algorithm we have now is built on the HSC transient survey, we could start finding lensed quasars in the HSC transient survey right away. Given the depth and the covering area, we estimate that $1\pm1$ quad(s) lie(s) in the footprint of the HSC transient survey up to June 2017, based on the lensing rates in OM10 and the correction for the faint end (\textit{i}-band total brightness $m_\text{total} > 18.1 \text{ mag}$) in \citet{Adriano_gaia_des_01}. There is already one known quad lying in this footprint \citep{Anguita_lens}. Therefore, we expect the recovery of this known lensed quasar and a possible discovery of a new quad when we apply our lens search algorithm on $\sim$141000 variable objects in the HSC transient survey. Most of the contaminants in the variable objects are very bright objects, such as bright variable stars or bright galaxies. (Chao et al. in prep.)\\

Moreover, by adapting our simulation pipeline for the time-varied lensed images, we could also apply the lens search algorithm to other cadenced surveys, such as the upcoming LSST. The LSST is expected to have similar image quality as the HSC transient survey, but with a much larger area coverage. Around 2000 lensed quasars with four multiple images are predicted to be detected in the LSST (OM10). Although the LSST pixel size ($0.2\arcsec$) is slightly larger than the HSC transient survey ($0.17\arcsec$), the effective region is expected to be still distinguishable between potential lensed quasar candidates and non-lensed variable sources, as long as we adopt a large value for $p_\text{thrs}\%$ ($p_\text{thrs} > 92.5\%$). Therefore, our lens search algorithm is still applicable for the LSST, if $p_\text{thrs}$ is sufficiently large. The application of single-epoch-based lens search in Sec.~\ref{sec:single_epo} in the LSST should mostly depend on the seeing of the epoch that is used for lens search. However, we still need the mock lenses simulated for the LSST to investigate in detail the impact on our lens search performance from their image quality and cadence strategy. We expect our lens search algorithm to skilfully capture the new lensed quasars in the ongoing and upcoming cadenced surveys.\\

\begin{acknowledgements}
We thank Brandon Kelly, Takashi Moriya, Hiroko Niikura, Masaomi Tanaka, and Atsunori Yonehara for useful discussions.  DCYC and SHS thank the Max Planck Society for support through the Max Planck Research Group for SHS. This research was supported in part by Perimeter Institute for Theoretical Physics. Research at Perimeter Institute is supported by the Government of Canada through the Department of Innovation, Science and Economic Development and by the Province of Ontario through the Ministry of Research, Innovation and Science. JHHC acknowledges support from the Swiss National Science Foundation (SNSF). This work was supported by JSPS KAKENHI Grant Numbers JP15H05892 and JP18K03693. ATJ is supported in part by JSPS KAKENHI Grant Number JP17H02868. This research made use of Astropy,\footnote{http://www.astropy.org} a community-developed core Python package for Astronomy \citep{astropy:2013, astropy:2018}. The Hyper Suprime-Cam (HSC) collaboration includes the astronomical communities of Japan and Taiwan, and Princeton University. The HSC instrumentation and software were developed by the National Astronomical Observatory of Japan (NAOJ), the Kavli Institute for the Physics and Mathematics of the Universe (Kavli IPMU), the University of Tokyo, the High Energy Accelerator Research Organization (KEK), the Academia Sinica Institute for Astronomy and Astrophysics in Taiwan (ASIAA), and Princeton University. Funding was contributed by the FIRST program from Japanese Cabinet Office, the Ministry of Education, Culture, Sports, Science and Technology (MEXT), the Japan Society for the Promotion of Science (JSPS), Japan Science and Technology Agency (JST), the Toray Science Foundation, NAOJ, Kavli IPMU, KEK, ASIAA, and Princeton University.
\end{acknowledgements}

\begin{appendix}
\section{Sensible separation or sensible time delay}
\begin{figure}
\centering
\includegraphics[width=\columnwidth]{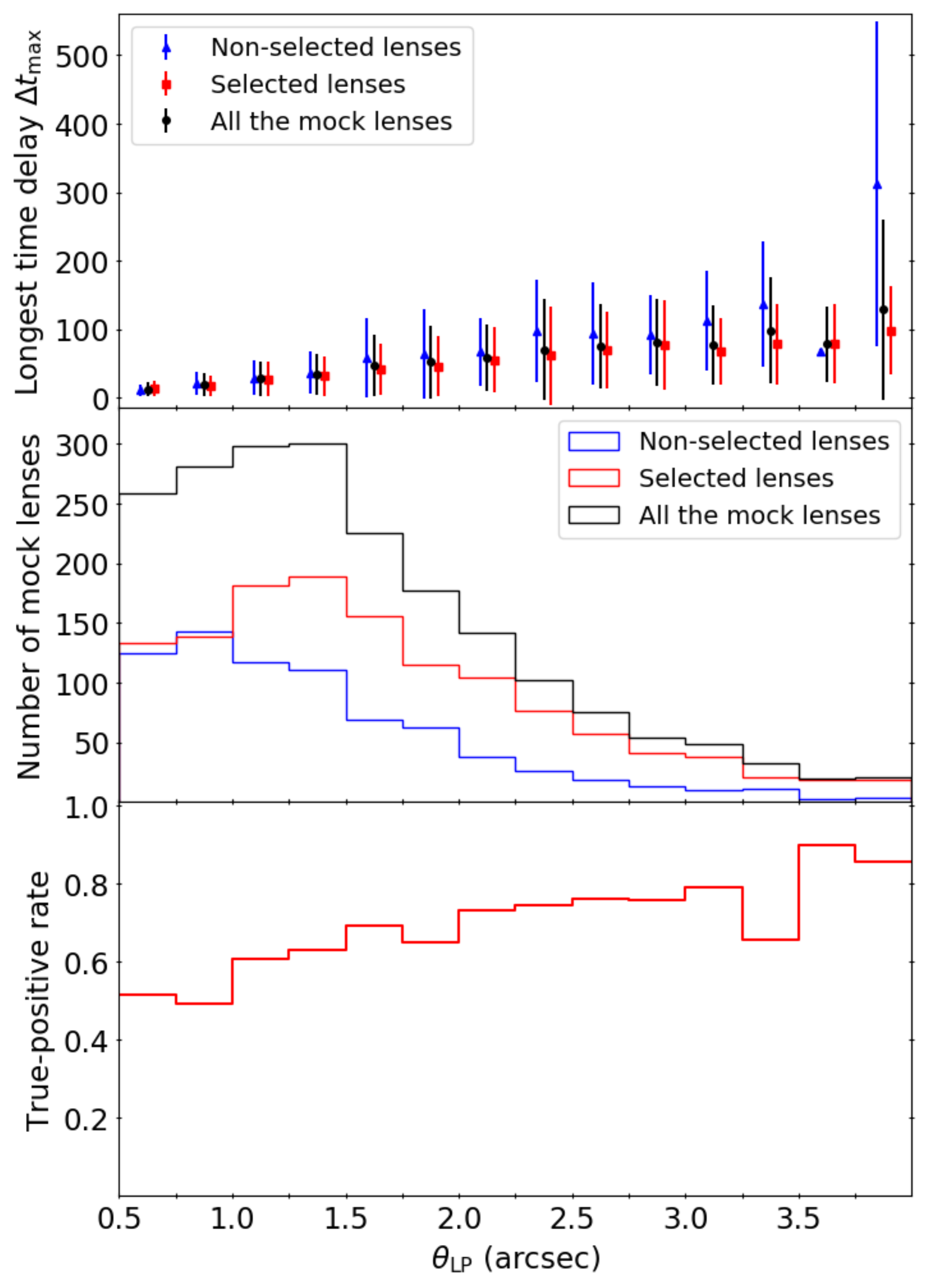}
\caption{Distribution of the longest time delay $\Delta t_\text{max}$ for the non-selected lensed quasars and the selected lensed quasars. The top panel shows the mean values of $\Delta t_\text{max}$ with $1$ standard deviation as the error bars from the lensed quasars in each $\theta_\text{LP}$ bin. The middle panel shows the number of non-selected lensed quasars, the number of selected lensed quasars, and the number of total mock lenses in each $\theta_\text{LP}$ bin. The bottom panel shows the TPR in each $\theta_\text{LP}$ bin. The sensitivity of the lens search algorithm mainly comes from the separation ($\theta_\text{LP}$), not the time delay ($\Delta t_\text{max}$).}
\label{fig:td_sep}
\end{figure}

We examine whether our lens search sensitivity has a dependence on the lensed quasar time delay by checking the longest time delay, $\Delta t_\text{max}$, in each lensed quasar. Fig.~\ref{fig:td_sep} shows the distribution of $\Delta t_\text{max}$ in each $\theta_\text{LP}$ bin. The top panel in Fig.~\ref{fig:td_sep} shows the mean value of $\Delta t_\text{max}$ with $1 \sigma$ standard deviation as the error bar from the lensed quasars in each $\theta_\text{LP}$ bin. The top panel in Fig.~\ref{fig:td_sep} shows that both the group of non-selected lensed quasars and the group of selected lensed quasars by our lens search method ($p_\text{thrs}\%=95\%$, $N_\text{thrs}=9$) have similar distribution of $\Delta t_\text{max}$ as all the lensed quasars, indicating that the lens search sensitivity has no preference on the lensed quasar time delay. On the other hand, the middle panel in Fig.~\ref{fig:td_sep} shows again that the number of selected lensed quasars decreases as the $\theta_\text{LP}$ decreases, while the number of non-selected lensed quasars increases as the $\theta_\text{LP}$ decreases. Therefore, our lens search algorithm is sensitive to the lensed quasar separation not to the lensed quasar time delay.\\
\section{Sensitivity to total (unresolved) brightness}
\begin{figure}
\centering
\includegraphics[width=\columnwidth]{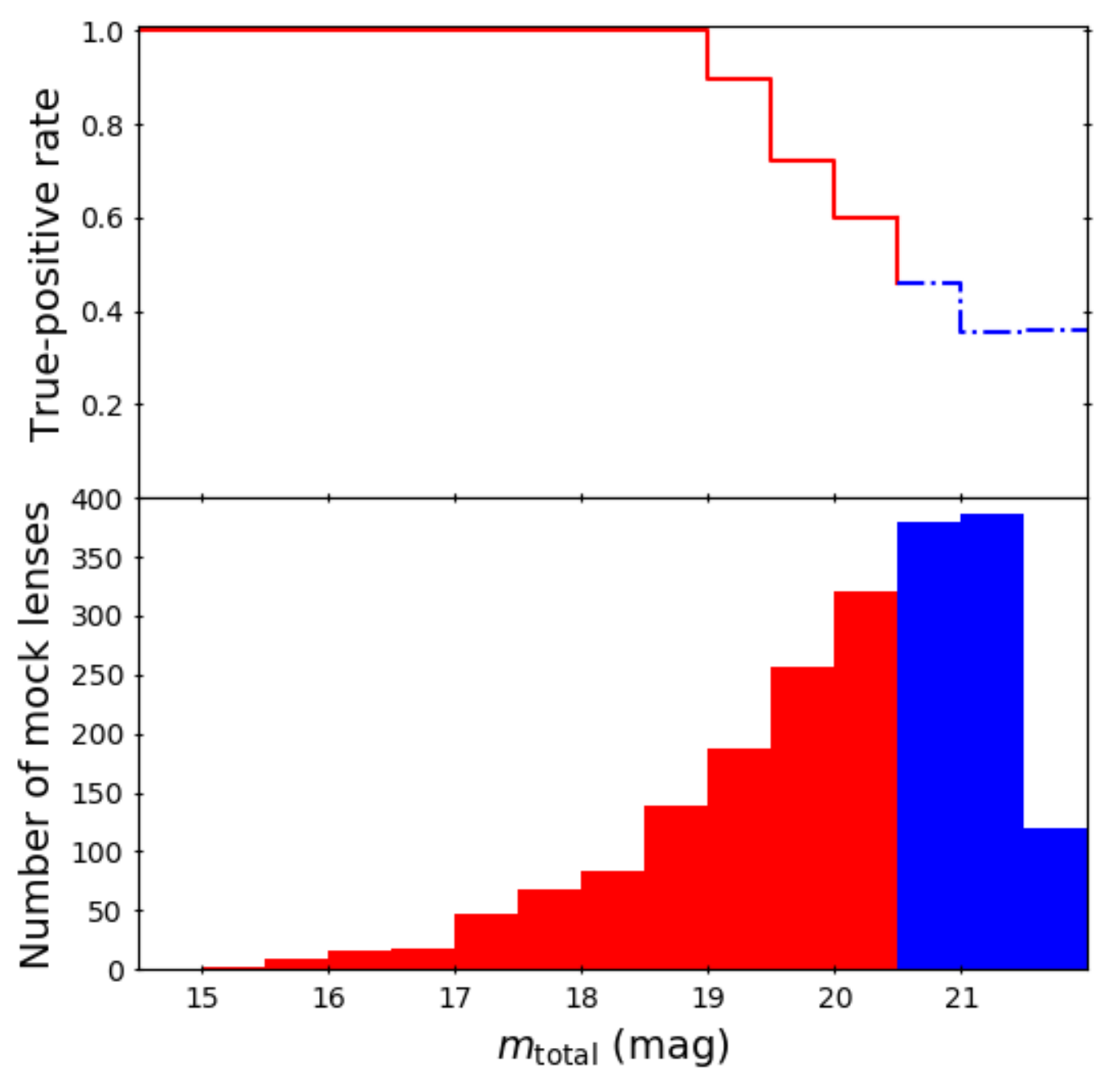}
\caption{Sensitivity to the lensed quasar total brightness $m_\text{total}$ of the lens search method based only on spatial extent, when $p_\text{thrs}\%=95\%$ and $N_\text{thrs}=9$. The lens search algorithm could capture all the lensed quasars with $m_\text{total}$ 19.0 mag.}
\label{fig:sensitivity_tot}
\end{figure}

\begin{figure}
\centering
\includegraphics[width=\columnwidth]{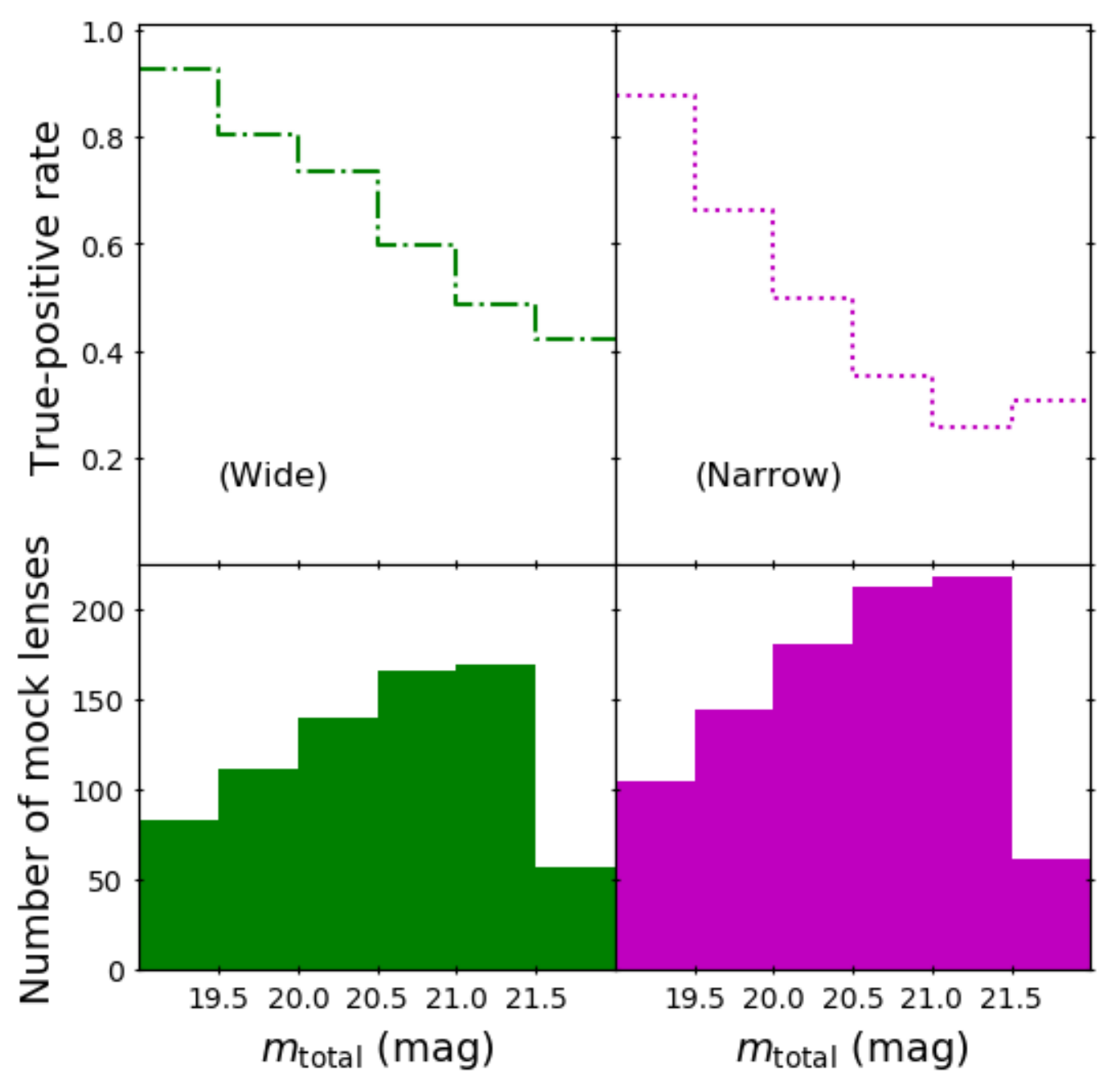}
\caption{Sensitivity to the lensed quasar total brightness $m_\text{total}$ for the lensed quasars fainter than 19.0 mag. Wide: The group of lensed quasars with the largest separation among the pairs $\theta_\text{LP} > 1.5\arcsec$. Narrow: The group of lensed quasars with $0.5\arcsec < \theta_\text{LP} \leq 1.5\arcsec$.}
\label{fig:sensitivity_faint}
\end{figure}

For a more realistic study, we investigate the lens search sensitivity to the lensed quasars total brightness, $m_\text{total}$ (\textit{i}-band). Fig.~\ref{fig:sensitivity_tot} shows the sensitivity to $m_\text{total}$, when applying the lens search method based only on spatial extent at $(p_\text{thrs}\%, N_\text{thrs})=(95\%, 9)$ (Sec.~\ref{sec:extendedness_only} and Sec.~\ref{sec:performance_basic}). As shown in Fig.~\ref{fig:sensitivity_tot}, regardless of the separation, our lens search algorithm could capture all the lensed quasars with $m_\text{total} < 19.0 \text{ mag}$. We further examine the sensitivity for the lensed quasars with $m_\text{total} > 19.0 \text{ mag}$ in Fig.~\ref{fig:sensitivity_faint}. Fig.~\ref{fig:sensitivity_faint} shows that our lens search algorithm can still identify more than 70\% of the lensed quasars with $19.0 \text{ mag} < m_\text{total} < 20.5 \text{ mag}$ when $\theta_\text{LP} > 1.5\arcsec$ (the largest separation among the pairs in one lens system),  while the lens search performance is relatively poor for the lensed quasars with $m_\text{total} > 19.5 \text{ mag}$ when $\theta_\text{LP} \leq 1.5\arcsec$. 

\end{appendix}

\bibliographystyle{aa} 
\bibliography{lens_search_time_variability.bib}

\end{document}